\documentclass{JHEP3} % 10pt is ignored!

\usepackage{epsfig,multicol,bbm,amsmath}
\usepackage{graphicx}
\usepackage{amssymb}
\newcommand{\bea}{\begin{eqnarray}}
\newcommand{\eea}{\end{eqnarray}\noindent}
\newcommand{\nn}{\nonumber \\}
\newcommand{\colH}[1]{#1}

\newcommand{\tr}{\text{tr}}

\newcommand{\Tr}{\text{Tr}}
\newcommand{\fss}[1]{#1\!\!\!/}   % Feynman slash short
\newcommand{\fsV}[1]{#1\!\!\!\!\!\;\!/\;} % and longer
\newcommand{\Eqref}[1]{Eq.~\eqref{#1}}

\newcommand{\Nf}{N_{\mathrm{f}}}

\newcommand{\yb}{\bar{\psi}}

\newcommand{\cm}{\text{CM}}
\newcommand{\cp}{\text{CP}}
\newcommand{\xcp}{x_\cp}
\newcommand{\sech}{\mathrm{sech}}

\graphicspath{{fig/}}

\title{Worldline Monte Carlo for fermion models at large $\Nf$}

\author{Gerald Dunne \\
        Department of Physics,  University of Connecticut,
        Storrs, CT 06269-3046, USA \\
        E-mail: \email{dunne@phys.uconn.edu}}

\author{Holger Gies \\
        Theoretisch-Physikalisches Institut,
        Friedrich-Schiller-Universit\"at Jena, D-07743 Jena\\
        \& Institut f\"ur Theoretische Physik, Universit\"at  Heidelberg
        D-69120 Heidelberg, Germany  \\
        E-mail: \email{holger.gies@uni-jena.de}}

\author{Klaus Klingm\"uller \\
        Institut f\"ur Theoretische Physik E, RWTH Aachen University,
        D-52056 Aachen\\ 
        \& Institut f\"ur Theoretische Physik, Universit\"at  Heidelberg
        D-69120 Heidelberg, Germany  \\
        E-mail: \email{klingmueller@physik.rwth-aachen.de}}

\author{Kurt Langfeld \\
        School of Maths \& Stats, University of Plymouth,
        Plymouth, PL4 8AA, England \\
        E-mail: \email{Kurt.Langeld@plymouth.ac.uk}}

\received{} 		%%
\accepted{}		%% These are for published papers.

\preprint{}		% OR: \preprint{Aaaa/Mm/Yy\\Aaa-aa/Nnnnnn}
\abstract{Strongly-coupled fermionic systems can support a variety of
  low-energy phenomena, giving rise to collective condensation,
  symmetry breaking and a rich phase structure. We explore the
  potential of worldline Monte Carlo methods for analyzing the
  effective action of fermionic systems at large flavor number $\Nf$,
  using the Gross-Neveu model as an example. Since the worldline
  Monte Carlo approach does not require a discretized spacetime,
  fermion doubling problems are absent, and chiral symmetry can
  manifestly be maintained. As a particular advantage, fluctuations in
  general inhomogeneous condensates can conveniently be dealt with
  analytically or numerically, while the renormalization can always be
  uniquely performed analytically. We also critically examine the
  limitations of a straightforward implementation of the algorithms,
  identifying potential convergence problems in the presence of
  fermionic zero modes as well as \colH{in} the high-density region. }


\begin{document}

\section{Introduction}

{The elaboration of the QCD phase diagram as a function of temperature
and baryon density represents a great challenge in modern particle and nuclear physics.
It has been studied experimentally at the RHIC collider at Brookhaven and at CERN.
It has also attracted intense theoretical  investigation using
computer simulations in the context of lattice gauge theory and random matrix
theory~\cite{Stephanov:2007fk}. A significant computational obstacle is the so-called
sign problem, for which a number of  approximate techniques have been developed:
(i) a Taylor expansion with respect to the
baryon chemical potential~\cite{Choe:2002mt,Allton:2002zi,Ejiri:2003dc}; (ii)
imaginary values for the chemical potential
$\mu $~\cite{Alford:1998sd,de Forcrand:2002ci,D'Elia:2002gd}; (iii)
overlap enhancing techniques~\cite{Fodor:2001au}.
Despite these impressive successes, our knowledge of the QCD phase diagram
is still rather limited. Other approaches include the study of QCD-like theories
such as 2-colour QCD~\cite{Hands:1999md,Kogut:2002cm,Hands:2006ve}, and
 high density quark matter  in a colour superconducting
state~\cite{Bailin:1983bm,Alford:1997zt,Schafer:1999jg,Pisarski:1999tv,Alford:2007xm}.
Of particular interest is the identification of inhomogeneous phases, analogous to the LOFF
states of condensed matter systems \cite{Alford:2000ze,Rajagopal:2000wf,Bowers:2002xr,Alford:2007xm,casalbuoni}. This idea has been studied deeply for Skyrme models \cite{klebanov,goldhaber,jackson,manton}, and for four-fermion models \cite{Rosenstein:1990nm,hands,casalbuoni}. Lower dimensional models also provide insight into chiral dynamics, and
early lattice studies of lower-dimensional models such as the Gross
Neveu model in the limit of many flavours have shown that inhomogeneous
background fields are key to understand the dense
phase~\cite{Karsch:1986hm}.  Recently, this model has
attracted a lot of interest since it was found analytically that
the high-density state of fermion matter forms an inhomogeneous
``baryon crystal''~\cite{Thies:2003br,Thies:2006ti,Schnetz:2004vr,Basar:2008im}.  This analytic crystalline phase has
 been confirmed by a lattice analysis \cite{deForcrand:2006zz}, and such studies have led to several
new insights concerning inhomogeneous phases \cite{bringoltz,Nickel:2008ng}.

Good chiral properties of the fermion action is of central importance
for an investigation of quark matter at intermediate densities, since
the high-density transition is driven by chiral dynamics.
Unfortunately, lattice fermion actions necessarily suffer from the
fermion doubling problem as firstly pointed out by
Nielsen and Ninomiya~\cite{Nielsen:1980rz}. Nowadays, staggered
fermions~\cite{Kogut:1974ag}, domain wall fermions~\cite{Kaplan:1992bt}
or Neuberger fermions~\cite{Neuberger:1997fp}, which are an explicit
realization of the Ginsparg-Wilson relation~\cite{Ginsparg:1981bj},
are widely used in numerical simulations. Despite  these advanced
formulations and great numerical efforts, it turns out still to be cumbersome to
achieve good chiral properties, such as a sufficiently small pion mass.

Here we propose to use the {\em worldline} approach to study interacting fermion
systems. Since the  worldline approach to the quark determinant does not use
a lattice discretization of space-time, it circumvents many of these
significant difficulties. Here, we will argue that the prospects
of the worldline approach are (i) exact chiral symmetry but yet
a fully numerical approach, (ii) analytic renormalization and (iii)
a clear description of Fermi surface effects.
The worldline method is a string-inspired approach to quantum field theory;
see~\cite{Schubert:2001he,Strassler:1992zr} for reviews.  It was further
developed into a viable tool for an efficient calculation of functional
determinants for arbitrary background fields~\cite{Gies:2001zp%,Gies:2001tj
}. Subsequently, {\em worldline numerics} has enjoyed a wide span of
applications ranging from the Casimir
effect~\cite{Gies:2003cv,Gies:2006bt%,Gies:2006cq,Gies:2006xe
} and fermion induced quantum interactions~\cite{Langfeld:2002vy} to the
description of pair production in inhomogeneous
fields~\cite{Dunne:2005sx,Gies:2005bz}. A worldline lattice formulation has
also been presented in \cite{Schmidt:2003bf}.

In this paper we apply a worldline Monte Carlo approach to interacting fermion
models, concentrating on the $(1+1)$-dimensional Gross-Neveu model
\cite{Gross:1974jv,dhn}.  We review these models in Section 2, and describe
the worldline representation of the effective action. The treatment of
inhomogeneous condensates is presented in Section 3, and the extension to
finite temperature and chemical potential is explained in Section 3, along
with results of the computations. Section 5 contains our conclusions, and
several appendices contain relevant technical details.

}

\section{Worldline representation of the effective action}

\subsection{Continuum formulation}

In this work, we mainly concentrate on the the $D=1+1$ dimensional Gross-Neveu
(GN) model defined by the Euclidean fermionic action  \cite{Gross:1974jv,dhn}
\begin{equation}
S_{\text{F}}= \int d^Dx
	\biggl(
		-\bar\psi
		\fss\partial \psi
		+ \frac{g^2}{2\Nf}  ( \yb \psi )^2
	\biggr), \label{eq:SF}
\end{equation}
where $\Nf$ denotes the number of flavors. The GN model has a discrete chiral
symmetry under $\psi\to \gamma_5 \psi$.  Our conventions for the Dirac algebra
in $D=1+1$ are $\gamma_1=i \sigma_3$, $\gamma_2=i\sigma_1$,
$\gamma_5=\sigma_2$, $\{ \gamma_\mu,\gamma_\nu \}=-2 \delta_{\mu\nu}$, but
most of the equations below generalize straightforwardly to higher
dimensions. With the aid of a Hubbard-Stratonovich transformation, we can
rewrite this model in terms of a mixed fermionic-bosonic action
\begin{equation}
S_{\text{FB}}= \int d^D x \bigg[ \frac{\Nf}{2 g^2}\sigma^2
    -\yb \fss{\partial} \psi - i \sigma
  \yb \psi  \bigg] \label{eq:SFB}
\end{equation}
with the bosonic scalar $\sigma(x)$ which transforms as $\sigma \to -\sigma$
under the discrete chiral symmetry. In this work, we actually confine
ourselves strictly to the large-$\Nf$ limit of this model which is equivalent
to a semiclassical approximation of the $\sigma$ sector. Integrating out the
remaining fermion fluctuations, results in the large-$\Nf$ effective action
\begin{equation}
  \Gamma[\sigma]= \int d^D x  \frac{\Nf}{2 g^2}  \sigma^2
  - \Nf\, \Tr \ln (-\fss{\partial} - i  \sigma ), \quad
  \Nf\to\infty. \label{eq:Gam0}
\end{equation}
The resulting Dirac operator $\mathcal D = i \fss{\partial} -\sigma$ has a
$\gamma_5$ hermiticity, $\mathcal{D}^\dagger=\gamma_5 \mathcal{D} \gamma_5$,
implying that the fermionic contribution to $\Gamma$ is real. In this case, we
can use the identity $\Tr \ln i\mathcal{D} = \frac{1}{2} \Tr
\ln \mathcal{D} \mathcal{D}^\dagger$ to obtain
\begin{equation}
\Gamma[\sigma]= \int d^D x  \frac{\Nf}{2 g^2} \sigma^2
 -\frac{\Nf}{2} \Tr \ln (-\partial^2 +
  \sigma^2  - i \fss{\partial} \sigma ), \quad \Nf\to\infty, \label{eq:Gam1}
\end{equation}
now involving a fluctuation operator of Laplace type.

\colH{A variety of fermionic models of a similar type exist. An example for a
  model with a continuous chiral symmetry is given by the Thirring model which
  is discussed in appendix \ref{app:thirring}. Here, we concentrate on }the
  Gross-Neveu model. The contribution from the fermion fluctuations to the
  effective action can be written in propertime form,
\begin{eqnarray}
\Delta\Gamma[\sigma]&:=&-\frac{\Nf}{2} \Tr \ln (-\partial^2 +
  \sigma^2 - i \fss{\partial} \sigma ) \label{eq:DelG0}\\
&=& \frac{\Nf}{2} \int_{1/\Lambda^2}^\infty \frac{dT}{T} \Tr\, e^{-(-\partial^2 +
  \sigma^2 - i \fss{\partial} \sigma ) T},\label{eq:DelG1}
\end{eqnarray}
where we have used a propertime UV cutoff $\Lambda$ for the sake of
definiteness; any other regularization scheme can similarly be
implemented. Interpreting the fluctuation operator in the exponent as
a quantum mechanical Hamiltonian $H$, we can introduce the path integral
representation of the propertime evolution operator $\exp(-H T)$,
\begin{equation}
\Delta\Gamma[\sigma]=\frac{\Nf}{2} \frac{1}{(4\pi)^{D/2}}\, d_\gamma
\int\limits_{1/\Lambda^2}^\infty \frac{dT}{T^{D/2+1}}  \!\!
\int\limits_{x(0)=x(T)} \!\!\mathcal{D} x \, e^{-\frac{1}{4} \int_0^T
  d\tau \, \dot{x}^2(\tau)} \, e^{- \int_0^T d\tau \sigma^2 } \Phi[\sigma],
\label{eq:DelG2}
\end{equation}
containing the spin factor
\begin{equation}
\Phi[\sigma]=\frac{1}{d_\gamma}\, \tr_\gamma \; {\cal P} \, \hbox{exp} \left(
  i \, \int
 _0^T d\tau \; \left( \fss{\partial} \sigma \, \right) \, \right) \; ,
\end{equation}
and $\sigma$ are functions of $x(\tau)$. The normalization of the
path integral in \Eqref{eq:DelG2} has been chosen such that $\int
\mathcal{D} x\, \exp(-\int \dot{x}^2/4)=1$. We also introduced the
dimensionality of the Dirac algebra $d_\gamma$, e.g.,
$d_\gamma=2^{D/2}$ in even dimensions, such that $\Phi[0]=1$.

\subsection{Propertime discretization}

For a given background of arbitrarily varying $\sigma$ and $V$ fields, the
analytic computation of the effective action $\Gamma$ is formidable. For a
numerical computation, the worldline representation is highly useful, since
the computation of the path integral in Eqs.~\eqref{eq:DelG2} or
\eqref{eq:DelG2T} can be done efficiently with Monte Carlo methods. Most
importantly, the algorithm can be formulated for any background $\sigma$ in
full generality.

An estimate for the path integral over an operator $\mathcal{O}[x]$ is
given by an average over a finite ensemble of closed worldlines
(``loops'') $x_\ell(\tau)$, where $\ell=1,\dots,n_{\text{L}}$.
\begin{equation}
\int\limits_{x(0)=x(T)} \!\!\mathcal{D} x \,\mathcal{O}[x]\,  e^{-\frac{1}{4} \int_0^T
  d\tau \, \dot{x}^2(\tau)} \equiv \langle \mathcal{O}[x] \rangle
  \simeq \frac{1}{n_{\text{L}}} \sum_{\ell=1}^{n_{\text{L}}}
  \mathcal{O}[x_\ell].
\label{eq:nL}
\end{equation}
Since the path integral is over all spacetime, it can be decomposed
into a spacetime integral over a ``center of mass'' (CM) and a path
integral over worldlines fixed to this common center of mass,
\begin{equation}
\langle \mathcal{O}[x] \rangle= \int d^D x_{\text{CM}} \langle
\mathcal{O}[x] \rangle_{x_{\text{CM}}}
\end{equation}
This latter ensemble of mass-centered worldlines $x_\ell$ is generated
according to a Gau\ss ian velocity distribution
\begin{equation}
P_{\text{CM}}[x]= \delta\left( x_{\text{CM}} - \int_0^T d\tau x(\tau) \right) \,
  e^{-\frac{1}{4} \int_0^T   d\tau \, \dot{x}^2(\tau)},
\end{equation}
where the $\delta$ function implements the center-of-mass
constraint. The center-of-mass integration relates the effective
action with a corresponding action density, i.e., the Lagrangian,
$\Gamma = \int d^D x_{\text{CM}} \mathcal{L}(x_{\text{CM}})$, but this
decomposition of the worldlines is not unique. Alternatively, the
worldlines in the ensemble can be fixed to have one common point (CP), say
at $\tau=0$,
\begin{equation}
P_{\text{CP}}[x]= \delta\left( x_{\text{CP}} - x(\tau=0) \right) \,
  e^{-\frac{1}{4} \int_0^T   d\tau \, \dot{x}^2(\tau)},
\end{equation}
such that $\langle \mathcal{O}[x] \rangle= \int d^D x_{\text{CP}}
\langle \mathcal{O}[x] \rangle_{x_{\text{CP}}}$.  The associated
Lagrangian \colH{$\mathcal{L}$}, $\Gamma = \int d^D x_{\text{CP}}
\widetilde{\mathcal{L}}(x_{\text{CP}})$, is not identical to the one
related to the center-of-mass definition, but agrees with it up to a
total derivative term \cite{Schubert:2001he}.  Below, we use both
prescriptions: whereas the CM worldlines have some computational
advantages due to the constraint $\int x(\tau)=0$, analytical results
are often related to the representation involving CP worldlines.

The finite ensemble of worldline loops still has an infinite set of degrees of
freedom. We reduce these to finitely many by a discretization of the
propertime coordinate $\tau$, resulting in an approximation of each worldline
by a finite set of $N$ points per loop, $x_i:= x(\tau_i)$, \colH{$\tau_i=T
  i/N$}, $i=0,\dots, N$, where the points $x_{i=0}\equiv x_{i=N}$ are
identified.  Replacing the propertime derivative in the worldline action by a
nearest-neighbor difference, $\dot{x}(\tau_i) \to \colH{N (x_i-x_{i-1})/T}$,
leads to a Gau\ss ian action for the $x_i$ which can be diagonalized including
the $\delta$ constraint. This allows for an ab-initio generation of a random
worldline distribution; for concrete algorithms, see
\cite{Gies:2003cv,Gies:2005sb}. A discretization in terms of Fourier modes is
also possible \colH{and will be used in Sect. \ref{sec:HMC}; it} shows a
slightly slower convergence towards the propertime continuum limit
\cite{Gies:2003cv}.

Whereas the finiteness of the number of worldlines introduces an error
which is measurable by the statistical error of the expectation
values, the propertime discretization leads to a systematic error
which has to be studied by approaching the continuum limit, i.e., by
increasing the value of $N$.

It should be stressed that only the auxiliary propertime is
discretized, whereas the spacetime coordinates remain continuous,
$x_i\in \mathbbm{R}^{D}$ (at least to machine accuracy). Therefore,
there is neither a fermion doubling problem nor is the algorithm
restricted to certain values of the flavor number $\Nf$; in fact,
$\Nf$ can even be a continuous number.

Another advantage of the worldline numerical approach is given by the fact
that chiral symmetries can manifestly be preserved. For instance in the case
of the Thirring model {
[discussed in more detail in Appendix \ref{app:thirring}]}, the SU($\Nf$)$_{\text{R}} \times$
SU($\Nf$)$_{\text{L}}$ remains unaffected also in the discretized version of
the worldline approach. This is because the derivation of the worldline
representation is completely done in the continuum, each step preserving the
chiral properties of the Dirac operator. The final worldline expression
\eqref{eq:DelG2T} hence is manifestly chirally invariant. Discretizing
the propertime does not spoil this invariance; moreover, spacetime remains
continuous even after propertime discretization. Even if we decided to discretize
the worldline integrals by constraining them
to form chains of links on a spacetime lattice, chiral symmetry would not be
violated, since such a discretization would not correspond to a lattice Dirac
operator.   We conclude that the numerical worldline Monte Carlo estimate for the
effective action respects chiral invariance, has no doubling problem and can
be used for any number of flavors $\Nf$.

\subsection{Renormalization}

Renormalization of these fermionic models in the large-$\Nf$ limit can be done
in the standard fashion, using, e.g., Coleman-Weinberg renormalization
conditions to fix physical parameters. The large-$\Nf$ limit even facilitates
renormalization in any dimension $D$, requiring, of course, an increasing
number of counter terms corresponding to physical parameters for increasing
$D$.\footnote{Beyond the large-$\Nf$ limit, the models are perturbatively
  renormalizable in $D=1+1$. For instance in $D=2+1$, they are even
  nonperturbatively renormalizable owing to the occurrence of a non-Gau\ss ian
  fixed point, as can be proven, e.g., to all order in a $1/\Nf$ expansion
  \cite{Gawedzki:1985ed,Rosenstein:1990nm}.}

In practice, however, renormalization can become an issue if the computation
is done numerically, since divergent pieces have to be separated from the
physical quantities in a well-controlled manner. Therefore, we briefly outline
the renormalization process in the following in order to demonstrate that the
renormalization can be done analytically first in our formalism, such that
numerics has to deal only with the remaining finite parts. For simplicity, we
confine ourselves to $D=1+1$ in the Gross-Neveu model but the same strategy
can be applied also in the case of other fermionic models.

The unrenormalized large-$\Nf$ effective action in the Gross-Neveu model reads
\begin{equation}
\Gamma[\sigma]= \int d^2 x  \frac{\Nf}{2 g_\Lambda^2} \sigma^2
+ \frac{\Nf}{(4\pi)}
\int\limits_{1/\Lambda^2}^\infty \frac{dT}{T^{2}} \int d^2 x
\langle  e^{- \int_0^T d\tau \sigma^2 } \Phi[\sigma] -1  \rangle_x,
\label{eq:Gunren}
\end{equation}
where we have already included the free-field subtraction, such that
$\Gamma[\sigma=0]=0$, and labeled the bare coupling with a subscript
$\Lambda$. A logarithmic cutoff dependence is generated from the small-$T$
limit of the propertime integrand. The small-$T$ expansion of the worldline
expectation value, which corresponds to the heat-kernel expansion, yields
\begin{equation}
\langle  e^{- \int_0^T d\tau \sigma^2 } \Phi[\sigma] -1  \rangle_x =
-\sigma^2(x) T + \mathcal{O}(T^2),
\end{equation}
where only the term linear in $T$ supports a $\log \Lambda$ divergence. This
divergence corresponds to the local counter term proportional to the bare
action and can be absorbed by a coupling renormalization. We fix the physical
coupling at the renormalization scale $\mu$ by a renormalization condition of
Coleman-Weinberg type:
\begin{equation}
\left. \frac{\delta^2 \Gamma[\sigma]}{\delta \sigma^2} \right|_{\sigma=\mu} =
  \frac{\Nf}{g^2(\mu)}.
\end{equation}
Inserting \Eqref{eq:Gunren}, this relates the bare to the renormalized
coupling,
\begin{equation}
	\frac 1 {g(\mu)^2}
=
	\frac 1 {g_\Lambda^2} + \frac 1 {2\pi}
	\left( C + 2 + \ln\frac{\mu^2}{\Lambda^2} \right),
\end{equation}
where $C$ denotes the Euler-Mascheroni constant. Also the $\beta$ function can
be read off, $ \beta_{g^2} \equiv \mu \partial_\mu g(\mu)^2 = - \frac 1 \pi
g(\mu)^4$. The finite parts are specific to our propertime cutoff
regularization; of course, any other desired regularization method can also be
used, such that the final numerical results can be mapped to any given
regularization scheme.

The renormalization scale and the coupling can be traded for an RG invariant
mass scale, manifesting dimensional transmutation. A convenient choice is
given by the minimum $m$ of the effective action which is directly related to
the induced fermion mass,
\begin{equation}
\frac{\delta \Gamma }{ \delta \sigma} \bigg|_{\sigma=m,\Lambda\to\infty} = 0\quad
\Longrightarrow\quad
	m
	=
	\mu\, e^{ -\frac{\pi}{g(\mu)^2} + 1}.
\label{eq:defm}
 \end{equation}
Expressing the bare coupling in terms of the renormalized coupling and
successively in terms of the minimum $m$, leads us to the final renormalized
form of the effective action,
\begin{equation}
\Gamma[\sigma]= \frac{\Nf}{4\pi} \int d^2 x\left[-   \sigma^2
+
\int\limits_{0}^\infty \frac{dT}{T^{2}} \left(
\langle  e^{- \int_0^T d\tau \sigma^2 } \Phi[\sigma]  \rangle_x
- \frac{\sigma^2}{m^2} (e^{-m^2 T}-1) -1 \right)\right],
\label{eq:Gren}
\end{equation}
which holds for any given background $\sigma(x)$. {Note that the worldline
  expectation value that is to be computed numerically is finite. } Together
with the subtraction terms, also the propertime integral is finite in the
limit $\Lambda\to\infty$, which we have performed already in
\Eqref{eq:Gren}. {For constant $\sigma$, we have $\langle e^{- \int_0^T
    d\tau \sigma^2 } \Phi[\sigma] \rangle_x=e^{-\sigma^2 T}$, and the
  effective action can be calculated analytically},
\begin{equation}
	\Gamma[\sigma=\text{const.}] = \frac{\Nf}{4\pi}\int d^2x
	\ \left[
	\sigma^2 \left( \ln\frac{\sigma^2}{m^2} - 1 \right)
	\right], \label{eq:Sconstshift}
\end{equation}
exhibiting a nontrivial minimum at $\sigma=m$ giving rise to chiral
symmetry breaking and fermion mass generation.

For the present work, the representation \Eqref{eq:Gren} is not only
conceptually satisfactory, but also useful from a practical viewpoint,
since the minimum $m$ which sets the physical scale is under control
analytically, cf. \Eqref{eq:Sconstshift}. For the remainder of this
subsection, we briefly consider an alternative but equally valid
choice of the scale.

The disadvantage of the above scale fixing lies in the fact that $m$
is determined from the minimum of the full effective action, which
happens to be at $\sigma=$constant in the large-$\Nf$ limit. In the
general case, e.g., beyond large $\Nf$ or in a fully nonperturbative
calculation, the true minimum of $\Gamma$ may not be known (or at
least relatable to the renormalized coupling) beforehand. Still, a
suitable subtraction scale is required for the computation. An
alternative choice is, for instance, given by the average field
$\bar\sigma^2=\frac{1}{\Omega_2} \int d^2 x \sigma(x)^2$, where
$\Omega_2$ is the $1+1$ dimensional spacetime volume. With this
choice, the large-$\Nf$ renormalized effective action becomes
\begin{equation}
\Gamma[\sigma]= \frac{\Nf}{4\pi} \int d^2 x\left[ \bar{\sigma}^2
	 \left( \ln\frac{\bar{\sigma}^2}{m^2} - 1 \right)
 +
\int\limits_{0}^\infty \frac{dT}{T^{2}} \left(
\langle  e^{- \int_0^T d\tau \sigma^2 } \Phi[\sigma]  \rangle_x
-  e^{-\bar{\sigma}^2 T}\right)\right],
\label{eq:Gren2}
\end{equation}
where $m$ is still defined via \Eqref{eq:defm}. This choice of the subtraction
scale has been used in \cite{Langfeld:2007wh}.  Beyond large $\Nf$ or fully
nonperturbatively, the first log term will be replaced by a more complicated
effective action, but the fermion loop contribution can still be computed in
this manner and remains finite.

\section{Spatially inhomogeneous $\sigma$ condensates}
\label{inhomog}

\subsection{Worldline formalism}

A number of nontrivial tests of the preceding general worldline approach can
already be performed within the Gross-Neveu model in $D=1+1$ which we will
exclusively consider in the following. For the case of static but spatially
varying $\sigma$ condensates, a number of exact results can be derived, and
these serve as stringent  tests and controls of our general worldline
approach. Special attention will be paid to bound states, especially zero
modes and near-zero modes, {as these pose the greatest numerical challenges.}

For a static but spatially varying $\sigma(x)$, the renormalized action in the
$1+1$ dimensional Gross-Neveu model at large $\Nf$ boils down to
\begin{equation}
	\Gamma[\sigma]
	= \frac{\Nf}{4 \pi}
	\int d^2x\  \Biggl[
	- \sigma^2
	+ \int_0^\infty \frac{dT}{T^2}
	\biggl(
	\Bigl\langle
	e^{ - \int_0^T d\tau\ \sigma^2 }
	\cosh( \int_0^T d\tau\ \sigma' )
	\Bigr\rangle_x
	-
	\frac{\sigma^2}{m^2} \left( e^{-m^2T} - 1 \right) - 1
	\biggr)
	\Biggr],
\end{equation}
where the prime denotes the derivative with respect to the spatial coordinate,
$\sigma'=\partial \sigma/\partial x$.  It is convenient to combine exponential
and hyperbolic function in the expectation value by defining
\begin{equation}
V_\pm(x) := \sigma^2(x) \pm \sigma^\prime(x) \quad,
\label{potentials}
\end{equation}
and
\begin{equation}
 \quad
	\Gamma = \frac \Nf 2 \int d^2x\ (\mathcal L_+ + \mathcal L_-)\quad ,
\end{equation}
where
\begin{equation}
	4 \pi \mathcal L_\pm := -V_\pm + \int_0^\infty \frac{dT}{T^2}
	\left(
		\left\langle e^{-\int d\tau V_\pm} \right\rangle_x
		-\frac{V_\pm}{m^2} \left( e^{-m^2 T}-1 \right) - 1
	\right). \label{eq:Lpm}
\end{equation}
So far, we have not specified the details of the worldline ensemble.  For a
comparison with analytical results, common-point (CP) loops are particularly
useful, since then we can write the associated expectation value as a quantum
mechanical transition amplitude in imaginary time,
\begin{equation}
	\left\langle e^{-\int d\tau V_\pm} \right\rangle_{\xcp}
	=
	\frac{\langle x | e^{-T (-\partial_1^2 + V_\pm)} | x \rangle}
	{\langle x | e^{-T (-\partial_1^2)} | x \rangle}
	=
	\sqrt{4 \pi T} \langle x | e^{-TH_\pm} | x \rangle
	\equiv\sqrt{4 \pi T}\, K_\pm(x;T) \quad .
	\label{eq:EVcp}
\end{equation}
Here we have introduced the Hamiltonian $H_\pm = -\partial_1^2 +
V_\pm$, and the heat kernel $K_\pm(x;T)$, with $x\equiv\xcp$ being  the common
point. We have normalized the heat kernel such that it agrees with the
standard conventions of 1D quantum mechanical transition
amplitudes. {For example, with a homogeneous condensate $\sigma(x)=m$,
we have $V_\pm=m^2$, and
\begin{equation}
K_\pm(x; T)=\frac{e^{-m^2 T}}{\sqrt{4 \pi T}} \quad .
\label{eq:free-propagator}
\end{equation}
The transition amplitude, or heat kernel, in turn} can be written as a sum
over the eigen modes $\psi_n$ of the Hamiltonian $H_\pm$:
\begin{align}
	K_\pm(x;T)\equiv\langle x | e^{-TH_\pm} | x \rangle
	=\sum_n |\psi_n(x)|^2 e^{-TE_{\pm}^n},
	\label{eq:transampl}
\end{align}
where the eigenvalues of the Hamiltonian are denoted by $E_{\pm}^n$.  This
spectral decomposition fixes the large-$T$ behavior of the worldline
expectation value.  If the Hamiltonian $H_\pm$ has no zero mode, then all terms of
the sum are exponentially damped and the expectation value vanishes for large
$T$. By contrast, if a zero mode exists and the spectrum is discrete at its
lower end the transition amplitude
Eq.~(\ref{eq:transampl}) maps out the zero-mode shape $|\psi_0(x)^2|$ and the
expectation value increases proportional to $\sqrt T$ for large $T$,
\begin{equation}
	\left\langle e^{-\int d\tau V_\pm} \right\rangle
	\stackrel{T \to \infty}{\longrightarrow}
	\sqrt{4 \pi T} |\psi_0(x)|^2.
	\label{eq:zm_contrib}
\end{equation}
The case with a zero mode is numerically challenging, as is discussed in detail below.

In the following sections, we present some exact expressions for $\langle
  e^{-\int_0^T d\tau\, V_\pm}\rangle$ for certain inhomogeneous condensates
  $\sigma(x)$, and we then compare these exact expressions to the numerical
  worldline results. These inhomogeneous condensates are taken from known
  inhomogeneous solutions to the gap equation of the GN model
  \cite{dhn,Thies:2003br,Thies:2006ti}.  This serves as a precise test and control of our
numerical scheme.  These exact results are derived from the corresponding
exact resolvents, derived in Appendix~\ref{app:oldA} using the Gel'fand-Dik'ii
equation (\ref{gd1}, \ref{gd2}), by an inverse Laplace transform:
\bea
R_\pm(x; -\lambda)\equiv \langle x\, | \frac{1}{H_\pm +\lambda}\,| x\rangle
=\int_0^\infty dT\, e^{-\lambda T}  \langle x\, | e^{-H_\pm \, T}\, |\,
x\rangle\ .
\label{resolvent}
\eea
For example, with a homogeneous condensate $\sigma(x)=m$,  $V_\pm=m^2$, so the resolvent follows trivially from (\ref{gd1}). We find
$R_\pm(x; -\lambda)=\frac{1}{2\sqrt{m^2+\lambda}}$, which leads to the
familiar result (\ref{eq:free-propagator}) by an inverse Laplace transform.

\subsection{Single kink condensate}

Consider the inhomogeneous
kink-like condensate, studied by inverse scattering in \cite{dhn},
\bea
\sigma(x)=A\,\tanh(A\, x) .
\label{kink}
\eea
The associated effective Schr\"odinger potentials (\ref{potentials}) are
\bea
V_\pm =\begin{cases}
A^2\cr
A^2(1-2\,{\rm sech}^2(A\, x))
\end{cases}.
\label{potentials-kink}
\eea
The kink condensate $\sigma(x)$, and the corresponding $V_\pm(x)$, are
plotted in Fig.~\ref{fig:kink}. The potential $V_-(x)$ has an exact
zero mode, localized on the kink.
\FIGURE{
\centerline{\includegraphics[scale=0.6]{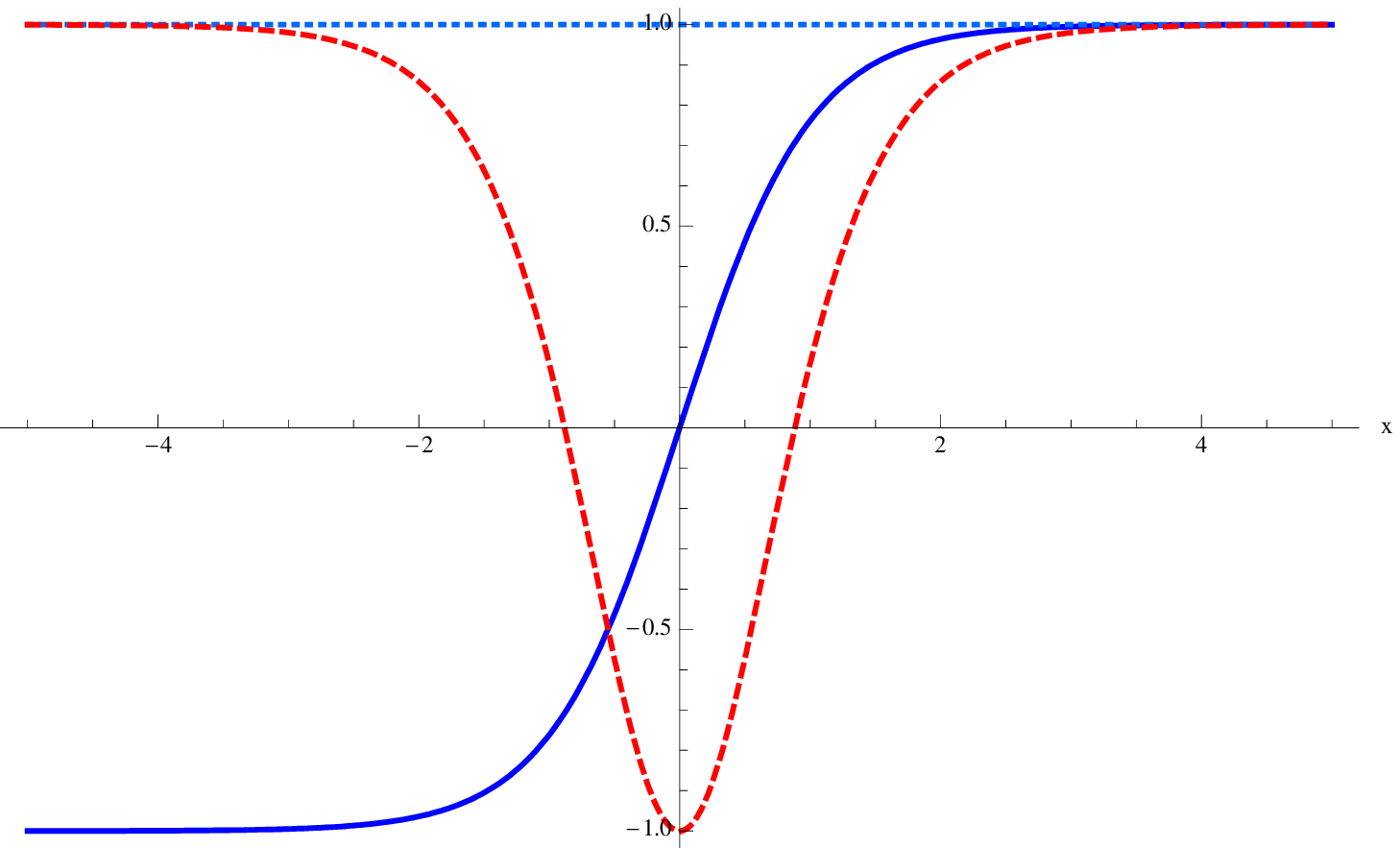}}
\caption{Plots of the kink condensate $\sigma(x)$ in
  \protect\Eqref{kink} [solid, dark blue, line], and the potentials $V_\pm(x)$ in
  \protect\Eqref{potentials-kink}. The potential $V_-$ [dashed, red, line]
  is localized on the kink, while $V_+$ [dotted, light blue, line] is
  constant. These plots are for $A=1$. }
\label{fig:kink}
}

The diagonal resolvents $R_\pm(x;-\lambda)$ are derived in Appendix~\ref{app:oldA},
and an inverse Laplace transform gives the corresponding exact heat kernels:
\bea
K_\pm(x; T)=
\begin{cases}
\frac{e^{-A^2\, T}}{\sqrt{4 \pi T}}\cr
\frac{e^{-A^2\, T}}{\sqrt{4 \pi T}}+\frac{A}{2}\, {\rm Erf}\left(A\sqrt{T}\right)\,
{\rm sech}^2(A\, x)
\end{cases},
\label{kink-propagator}
\eea
{where the error function is ${\rm
    Erf}(z)=\frac{2}{\sqrt{\pi}}\int_0^ze^{-t^2}dt$. } According to
Eqs.~\eqref{eq:transampl}, \eqref{eq:zm_contrib}, the zero mode of the $V_-$
potential is reflected in the large $T$ behavior of $K_-(x; T)$: $K_-(x;T\to
\infty)\to\frac{A}{2}\sech^2(Ax) \equiv |\psi_0(x)|^2$.

Capturing the zero-mode physics is indeed a precarious if not
pathological problem for the standard worldline algorithm discussed
above. Close to the origin, the potential $V_-$ is negative, see
Fig.~\ref{fig:kink}, and consequently the exponent in $e^{-\int_0^T
d\tau V_-(x(\tau))}$ is positive for certain loops.  Comparatively
small loops with many points close to the origin can exponentially
dominate the expectation value for large $T$, inducing an overlap
problem for the Monte Carlo estimate.  A similar overlap
problem has been studied in \cite{Gies:2005bz}.

According to \Eqref{eq:zm_contrib}, the zero mode leads to a $\sqrt{T}$
increase of the exact worldline expectation value. It is immediately clear
that a finite and discrete loop ensemble cannot capture this property for
$T\to\infty$: since the distance between two neighboring points of a
discretized loop scales with $\sqrt{T}$, any loop ensemble will eventually no
longer resolve the negative peak of $V_-$ for large $T$, losing the
information about the zero mode. Thus, for any finite, discrete loop ensemble
generated with respect to the free worldline action, the expectation value eventually
decreases exponentially in the large $T$ limit.

In Fig.~\ref{fig:overlap}, the analytic result for the expectation value
$\langle e^{-\int_0^T d\tau V_-(x(\tau))}\rangle$ at the origin, i.e., the
center of the kink, for $A=1$ is compared to the worldline numerical result
for two loop ensembles with different numbers of loops.  For small
propertimes, the results agree nicely. However, for large $T$~values, the
numerical results indeed decrease exponentially; the values obtained using a
smaller loop ensemble decrease more rapidly than the result computed with more
loops. In this large $T$~region, the statistical errors are much smaller than
the real error, which is typical for an overlap problem. Let us stress,
however, that the worldline estimate is still reliable for a much larger range
of propertime than the first- or second-order heat-kernel expansion, which is
an asymptotic small-$T$ expansion [see Appendix \ref{app:hke}] that fails for
propertimes larger than $\mathcal{O}(1)$. {These heat-kernel expansion
  approximations are also plotted in Fig.~\ref{fig:overlap}, and we see
  clearly that the worldline result is far superior at large propertime $T$. }
\FIGURE{
	\centering
	\includegraphics[width=0.59\columnwidth]{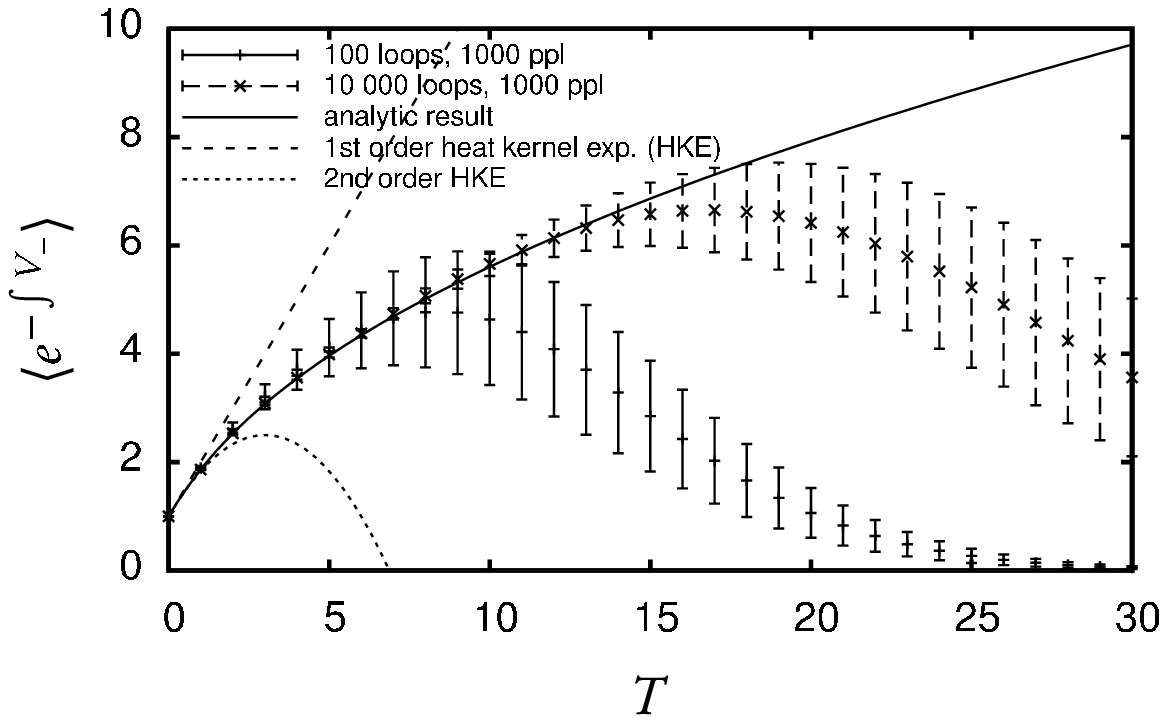}
	\includegraphics[width=0.39\columnwidth]{ShapeOfKinkZeroMode2}
	\caption{Left panel: the CP-loop expectation value at the centre of a
          kink potential with $A=1$ versus the propertime $T$.  For large
          propertimes, the analytic result is proportional to $\sqrt T$. This
          is rediscovered by worldline numerics only up to a certain
          propertime value, depending on the number of loops employed. Still,
          the worldline estimate is reliable over a much larger range of
          propertimes than the first- and second-order heat-kernel
          expansion. Right panel: zero-mode wave function of the kink with
          $A=1$ reconstructed from the worldline computation of the heat
          kernel at large $T$ using a CP loop ensemble.}
	\label{fig:overlap}
}

We conclude that the zero-mode-induced overlap problem inhibits a
straightforward calculation of the heat kernel's large $T$ behavior
and can only be shifted to larger $T$ values by using a larger loop
ensemble. In Subsect.~\ref{sec:HMC}, we demonstrate that the overlap problem
can be solved by generating an adapted ensemble with a Hybrid Monte Carlo
algorithm.

In the present kink case, the zero-mode physics can still be identified with
the standard free ensemble, since it is clearly separated from the remaining
spectrum. Already at finite but large $T$, the heat kernel is essentially
determined by the zero-mode contribution, with all other modes being
exponentially suppressed. The right panel of figure \ref{fig:overlap} shows
the zero-mode shape extracted from the heat kernel at a range of propertime
values $T$ of $\mathcal{O}(10)$ using 50000 CP loops consisting of $100$
points per loop. In this propertime range, the modulus of the zero-mode
wavefunction $\vert \psi _0 \vert ^2 $ can be reconstructed from
(\ref{eq:transampl}) by a fit. Comparison with the exact result reveals that a
rather rough discretisation of the loops already yields quite accurate
results.

\subsection{Kink-antikink pair condensate}

The single-kink condensate discussed in the previous subsection has an exact
zero mode for $V_-(x)$, which complicates the worldline numerics at large
$T$. {In this section, we probe this phenomenon more precisely by
  considering a condensate for which the bound state is shifted away from zero
  energy. } Consider the kink-antikink configuration:
\bea
\sigma(x)&=&A\left\{\coth(b)+\left[\tanh(A\, x)-\tanh(A\, x+b)\right]\right\} \nn
&=& A\left\{\left[\coth(b)-\tanh(b)\right]+\tanh(b)\tanh(A\, x)\tanh(A\, x+b)\right\}.
\label{kink-antikink}
\eea
which has the form of a kink-antikink pair, with finite separation
$b/A$. The corresponding potentials (\ref{potentials}) are
\bea
V_\pm &=&A^2
\left( \coth^2(b)
-2 \left\{
\begin{matrix}
{\rm sech}^2(A\, x+b) \\
{\rm sech}^2(A\, x)
\end{matrix}
\right\}
\right).
\label{potentials-kink-antikink}
\eea
The kink-antikink condensate $\sigma(x)$, and the corresponding
$V_\pm(x)$, are plotted in Figure \ref{fig:kink-antikink}.
The potentials in \Eqref{potentials-kink-antikink} are special in the sense
that they are reflectionless, and each has a {\it single} bound state, located
at $A^2/{\rm sinh}^2(b)$, and a continuum starting at $A^2\,{\rm coth}^2(b)$.
Note that in the limit $b\to\infty$, the antikink disappears to minus infinity
and we are left with the single kink configuration $\sigma(x)=A\, \tanh(A\,
x)$, as in \Eqref{kink}. To see this pictorially, compare the second plot in
Figure \ref{fig:kink-antikink}, in the vicinity of $x=0$, with Figure
\ref{fig:kink}. Correspondingly, the bound state at $A^2/{\rm sinh}^2(b)$
becomes a zero mode in this infinite separation limit.
\FIGURE[t]{
\centerline{\includegraphics[width=0.49\columnwidth]{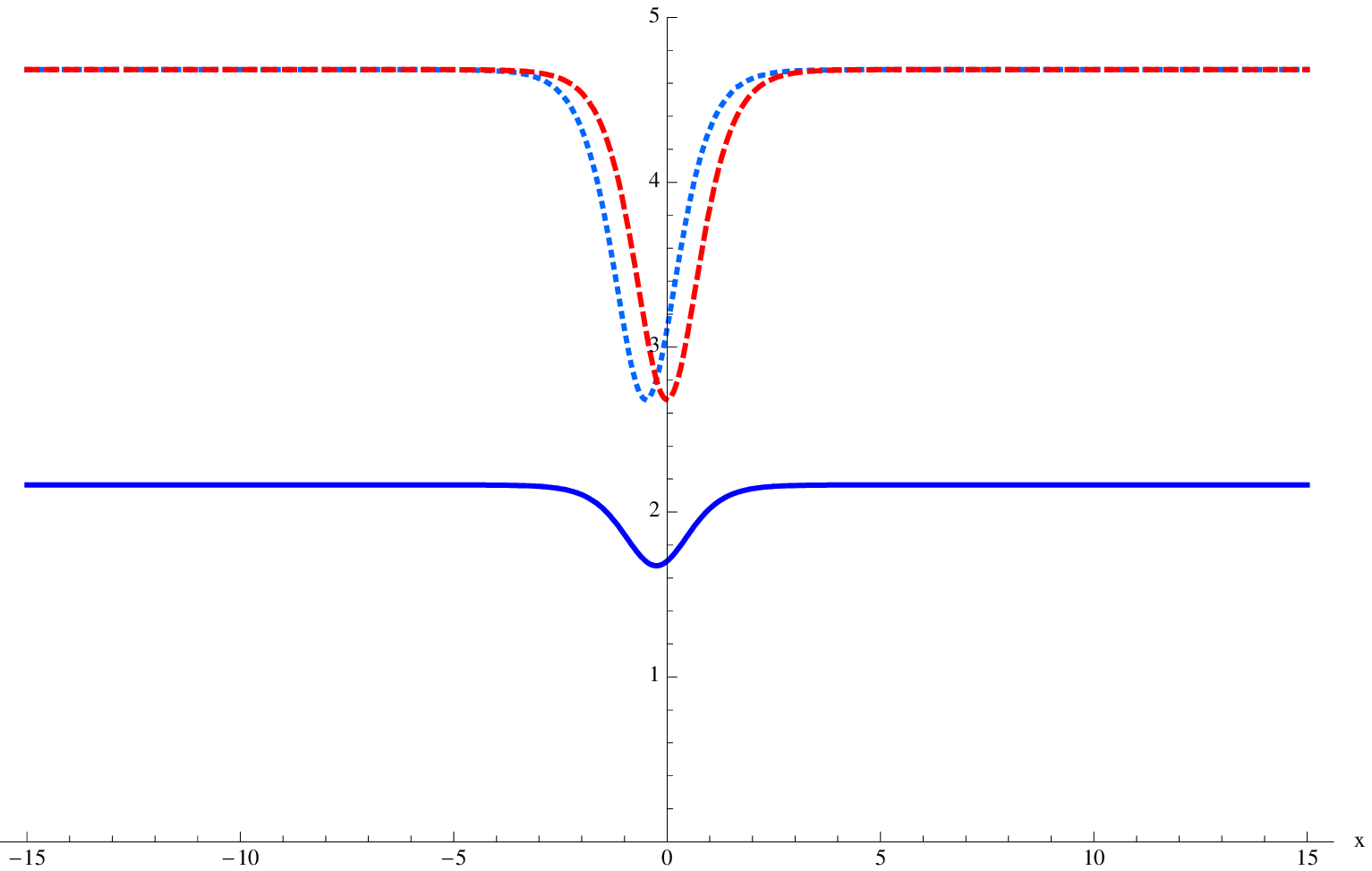}
\includegraphics[width=0.49\columnwidth]{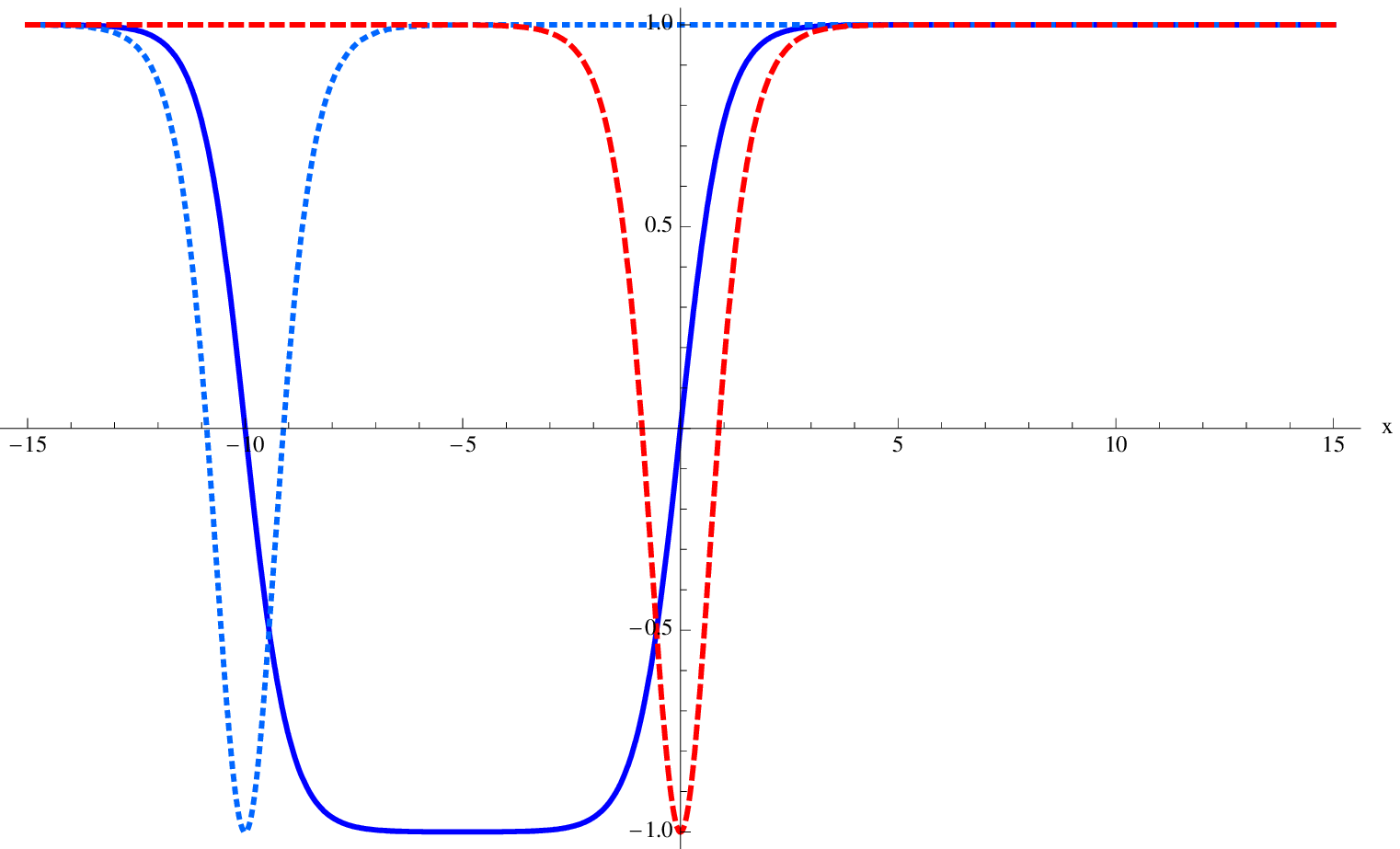}}
\caption{Plots of the kink condensate $\sigma(x)$ in
  \protect\Eqref{kink-antikink} [solid, dark blue, lines], and the potentials $V_\pm(x)$ in
  \protect\Eqref{potentials-kink-antikink}. The first plot is for $b=0.5$,
  while the second plot is for $b=10$. Both plots have $A=1$. Note that, in
  each case, the potential $V_-$ [dashed, red, line] is localized on the kink, while
  $V_+$ [dotted, light blue, line] is localized on the anti-kink. }
\label{fig:kink-antikink}
}

The resolvents $R_\pm(x; -\lambda)$ for the potentials $V_\pm$ are derived in
Appendix~\ref{app:oldA}, and
the heat kernels  follow from the inverse Laplace transform (\ref{resolvent}):
\bea
K_\pm(x; T)=\frac{e^{-A^2\, \coth^2(b)\, T}}{\sqrt{4 \pi T}}+\frac{A}{2}\,
e^{-A^2 T/\sinh^2(b)}\, {\rm Erf}\left(A\sqrt{T}\right)  \left\{
\begin{matrix}
{\rm sech}^2(A\, x+b) \\
{\rm sech}^2(A\, x)
\end{matrix}
\right\}.
\label{kkbar-propagator}
\eea
For finite $b$, the bound state at $A^2/{\rm sinh}^2(b)$ is reflected in the
exponential factor in the second term in (\ref{kkbar-propagator}). Thus, the
large-$T$ behaviour isolates the lowest energy mode, the bound state at
energy $E_{\text{b.s.}}$:
\begin{equation}
K_\pm(x; T)\to e^{-E_{\text{b.s.}}T} |\psi^{\text{b.s.}}_\pm(x)|^2\quad, \quad
T\to \infty.
\label{kkbar-limit}
\end{equation}
In the limit $b\to\infty$, this bound state becomes a zero mode, and the
heat kernels in (\ref{kkbar-propagator}) reduce to those of the single-kink
(\ref{kink-propagator}).  In the limit $b\to 0$, $A\to 0$, with
$A\,\coth b\equiv m$ fixed, $\sigma(x)\to m$, and $K_\pm(x; T)$ reduce
to the heat kernels (\ref{eq:free-propagator}) appropriate for a
homogeneous condensate.
For a comparison with worldline Monte Carlo results, we only consider the
challenging bound-state contribution contained, e.g., in $K_-(x;T)$ {[clearly
  $K_+(x; T)$ yields the analogous result for ${\mathcal L}_+$].}
This corresponds to the second term in \Eqref{kkbar-propagator}, yielding a
bound-state contribution to the action density, i.e., the Lagrangian,
\begin{equation}
  \mathcal{L}_{-,\text{b.s.}}(x)= \frac{1}{4\pi} \frac{A}{2} \sech^2(Ax) \int_0^\infty
  \frac{dT}{T}\, e^{-A^2 T/\sinh^2(b)}\,
  \text{Erf}(A\sqrt{T}) - \text{c.t}. \label{eq:Lminus1}
\end{equation}
{For illustration, we fix the parameter $b$ by $A\coth b=1$, so that the bound
  state energy is at $E_{\text{b.s.}}=1-A^2$. This choice minimizes the
  contribution of the continuum part of the spectrum to the action.} Including
explicit counter terms, satisfying Coleman-Weinberg renormalization
conditions, we obtain (setting $m=1$)
\begin{equation}
  \mathcal{L}_{-,\text{b.s.}}(x)= \frac{A}{4\pi} \, \sech^2(Ax) \int_0^\infty
  \frac{dT}{T^2} \left(\sqrt{\pi T}\,  e^{-(1-A^2) T}\,
  \text{Erf}(A\sqrt{T})
  + 2 A ( e^{-T} - 1)  \right) . \label{eq:Lminus2}
\end{equation}

The remaining free amplitude parameter $A$ determines the depth of the $V_-$
potential: {$A = 1$ corresponds to the single kink, with an exact zero
  mode, whereas $A = 0$ yields a constant, positive potential, with no bound
  state; intermediate values interpolate between both extremes.} If $A$ is
larger than $1 / \sqrt 2$, the potential becomes negative at the origin.  But
only for $A = 1$, there is a zero mode and the integrand in \Eqref{eq:Lminus2}
is proportional to $\sqrt T$ in the limit $T \to \infty$.

We have performed a worldline numerical computation for five potentials with
$A$ values between $1 / \sqrt 2$ and 1, using the standard free ensemble in
order to study the overlap problem quantitatively. The result corresponding to
the propertime integral in \Eqref{eq:Lminus2} is shown in Figure
\ref{fig:overlap2}.  {This analysis shows that, in general, the negative
  minimum of the potential is not a problem. Only for the largest $A$~value,
  $A = 0.995$, for which $E_{\bf b.s.}=0.00099975$, does the algorithm have a
  severe convergence problem. } In this case, the influence of the near-zero
mode becomes noticeable: the exponential factor in the integrand of
\Eqref{eq:Lminus2} decreases only weakly, and the integral is dominated by
large propertime values, at which the worldline numerical evaluation of the
expectation value suffers from the overlap problem, similar to that displayed
in Fig. \ref{fig:overlap}.
Whether or not a severe overlap problem occurs can be estimated from the
near-zero mode properties. A near-zero mode creates a $\sqrt{T}$ increase of
the worldline expectation value at intermediate $T$ values, being followed by
an exponential decrease at very large $T$ values. The crossover occurs at $T$
values $T\sim 1/E$, where $E$ corresponds to the near-zero mode's energy
level. For instance, for the above kink-antikink case with $A=0.9995$, the
crossover occurs at $T\simeq(1-A^2)^{-1}\sim 10^{3}$, implying that a reliable
estimate of the integrand up to values of $T\sim 10^3$ would be needed in
order to solve the overlap problem by brute force. For the case with
$A=0.942809$, the crossover occurs already at $T\simeq(1-A^2)^{-1}\sim 9$
which is well resolvable by the standard algorithm with 1000 loops or more,
cf. Fig.~\ref{fig:overlap}.
\FIGURE{
	\centering
	\includegraphics[width=.97\columnwidth]{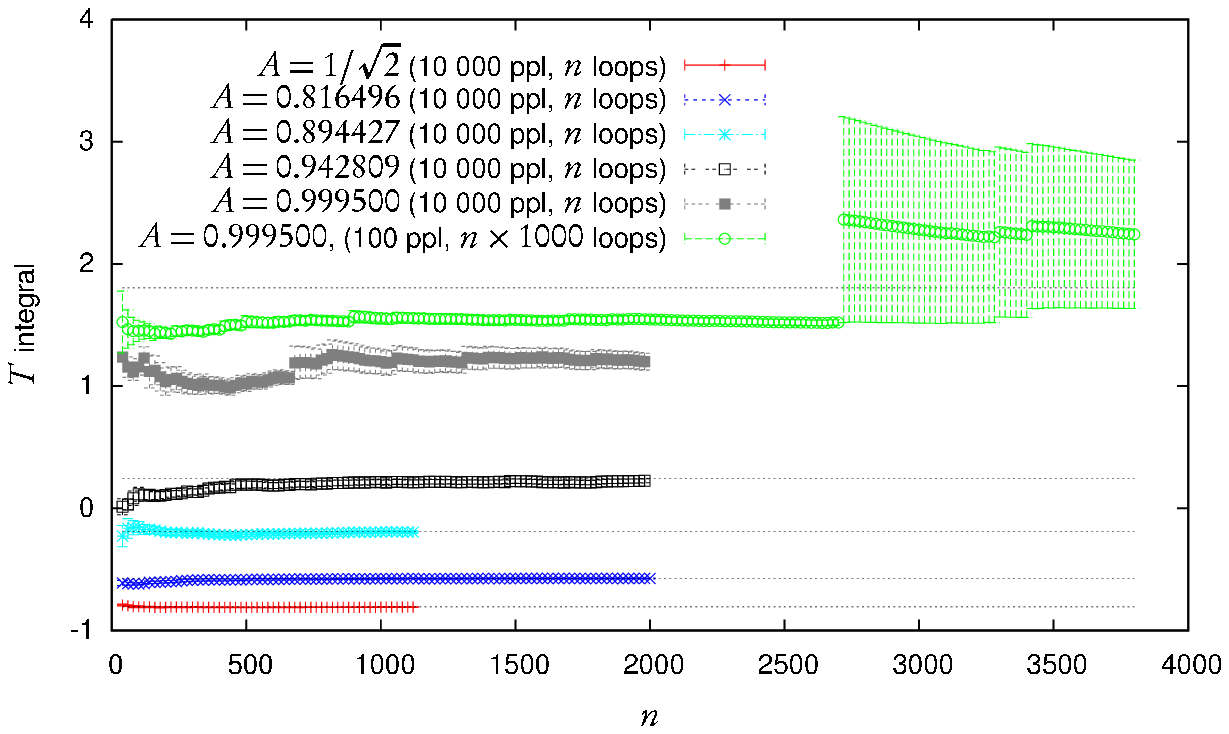}
	\caption{ Convergence behaviour of the worldline-numerical result for
          near-zero mode contributions. The plot shows the numerical values
          corresponding to the propertime integral \protect\Eqref{eq:Lminus2}
          versus the number of loops employed for different $A$~values. The
          topmost values (green circles) are for the same $A$~value as the
          values immediately below (solid grey squares), but with the number
          of loops multiplied by a factor of 1000.  The dotted horizontal
          lines represent the analytic values.  The error bars are the
          jackknife estimates and should provide reasonable values if the
          number of loops is larger than 1000.  Evidently, this is not the
          case for the largest $A$ value.  This indicates an overlap
          problem. At about 2~750~000 loops ($n$ = 2750), we observe a huge
          jump of the result, apparently caused by the occurrence of one
          single very small loop.  }
	\label{fig:overlap2}
}

\subsection{Periodic array of kink-antikink pairs}

Next, consider a crystalline condensate formed from a periodic array of
kink-antikink pairs:
\bea
\sigma(x)&=&A\left(\left[ Z(b; \nu)+\frac{{\rm cn}(b; \nu){\rm dn}(b; \nu)}{{\rm sn}(b; \nu)}\right]+\left[Z(A\, x; \nu)-Z(A\, x+b; \nu)\right]\right)\nn
&=&A\left(\frac{{\rm cn}(b; \nu){\rm dn}(b; \nu)}{{\rm sn}(b; \nu)}+\nu\, {\rm sn}(b; \nu)\, {\rm sn}(A\, x; \nu)\, {\rm sn}(A\, x+b; \nu)\right) \quad .
\label{kk-crystal}
\eea
Here $Z(x; \nu)$ is the Jacobi zeta function, {and sn, cn and dn the
  Jacobi elliptic functions, all with real elliptic parameter $0\leq\nu\leq 1$
  \cite{as,ww}. This type of inhomogeneous condensate is physically important
  in a number of contexts: it represents a bipolaron crystal in conducting
  polymers \cite{brazovskii}, and it characterizes a crystalline phase in the
  massive GN model \cite{Thies:2003br}.}  In the limit $\nu\to 1$ we recover
the kink-antikink configuration (\ref{kink-antikink}), since in this limit:
${\rm sn}(b;\nu)\to \tanh(b)$, $Z(b; \nu)\to \tanh(b)$, ${\rm cn}(b; \nu)\to
{\rm sech}(b)$, and ${\rm dn}(b; \nu)\to {\rm sech}(b)$. Note that we can
restrict to $0\leq b\leq {\bf K}(\nu)$, because of the periodicity of the
  Jacobi elliptic functions.  When $b={\bf K}(\nu)$, this condensate is
  important for soliton lattices in polymers \cite{brazovskii,horovitz}, for
  the Peierls model \cite{mertsching}, for periodic phases in superconductors
  \cite{buzdin}, and for the crystal phase of the massless GN model
  \cite{Thies:2003br}. In this case, discussed in more detail in the next
  subsection, the neighbouring antikinks lie at the very edge of the period,
and so in the infinite period limit [$\nu\to 1$], the antikinks disappear to
plus/minus infinity, and we are left with the single-kink configuration
(\ref{kink}).  Thus, the configuration (\ref{kk-crystal}) contains all the
other examples as special cases in a particular parametric limit.

The potentials (\ref{potentials}) corresponding to the periodic condensate (\ref{kk-crystal}) are
\bea
V_\pm=A^2\left( \frac{1}{{\rm sn}^2(b; \nu)}-1+\nu -2\nu  \left\{
\begin{matrix}
{\rm cn}^2(A\, x+b; \nu) \\
{\rm cn}^2(A\, x; \nu)
\end{matrix}
\right\}
\right) \quad .
\label{potentials-kk-crystal}
\eea
The kink-antikink crystal  condensate $\sigma(x)$, and the corresponding $V_\pm(x)$, are plotted in Figure \ref{fig:kk-crystal}.
\FIGURE[t]{
\centerline{\includegraphics[scale=0.49]{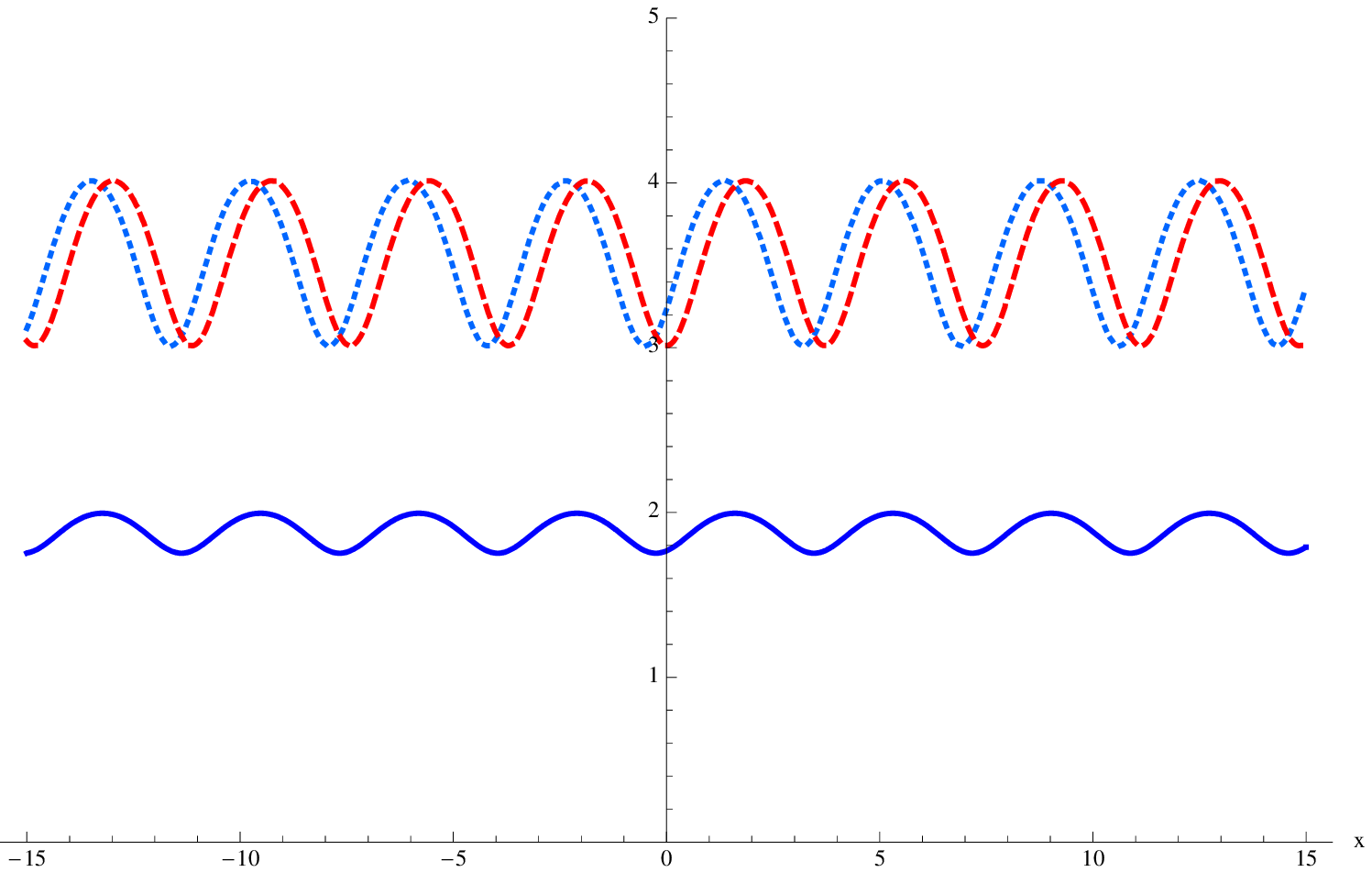}
\includegraphics[scale=0.49]{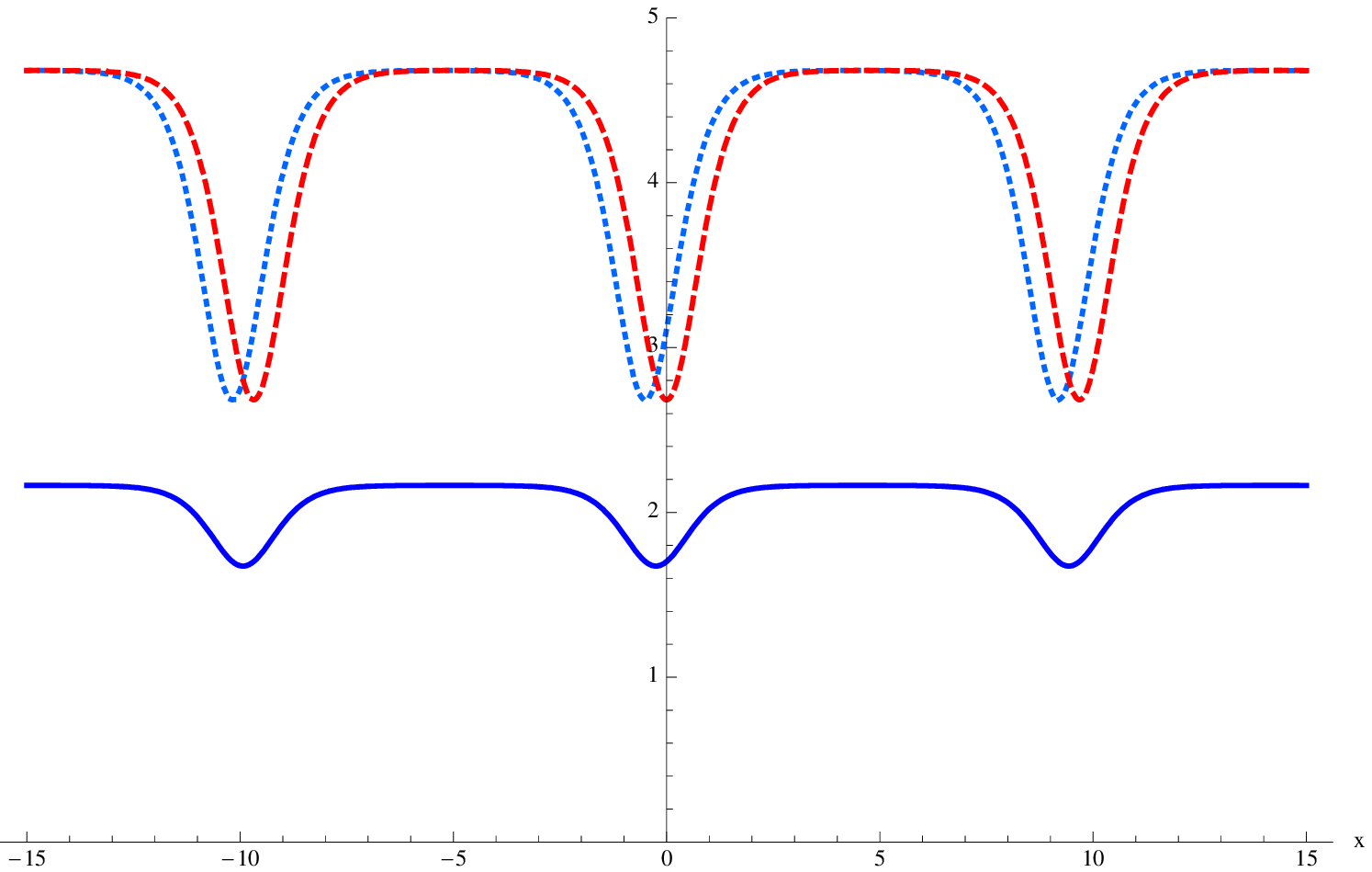}}
\caption{Plots of the periodic condensate $\sigma(x)$ in
  \protect\Eqref{kk-crystal} and of the periodic potentials
  $V_\pm(x)$ in \protect\Eqref{potentials-kk-crystal}. The first plot is for elliptic parameter $\nu=0.5$, for which ${\bf K}(0.5)=1.85$, while the second plot is for $\nu=0.999$, for which ${\bf K}(0.999)=4.84$. In each plot, $A=1$ and $b=0.5$.
The solid [dark blue] curves are the periodic condensates. Note that $V_-(x)$ [dashed, red, lines] are localized on the kink, while $V_+(x)$ [dotted, light blue, lines] are localized on the antikink. Notice that as $\nu\to 1$, the period diverges, and we recover the kink-antikink configuration: compare the second plot with the first plot in Figure \protect\ref{fig:kink-antikink}.}
\label{fig:kk-crystal}
}

These potentials are periodic functions, with period $2\,{\bf K}(\nu)/A$, and
are special in the sense that they possess only a {\it single} bound band --
they are the simplest example of the "finite-gap" potentials \cite{novikov},
which are the periodic analogue of reflectionless potentials.  $V_\pm(x)$
are the same function, displaced in $x$, and so they have precisely the same
band spectrum. In the language of supersymmetric quantum mechanics, they are
self-isospectral \cite{Dunne:1997ia}. The band spectrum is plotted in
Figure \ref{fig:kk-crystal-spectrum}. The band edges lie at $A^2\,{\rm
  cs}^2(b; \nu) $ and $A^2\,{\rm ds}^2(b; \nu)$, and with a continuum starting
at $A^2\,{\rm ns}^2(b; \nu)$.  As $\nu\to 1$, the period diverges, and the
bound band shrinks to a single bound state at $A^2/{\rm sinh}^2(b)$, which is
just the bound state of the reflectionless system associated with the single
kink-antikink pair.

\FIGURE[t]{
\centerline{\includegraphics[scale=0.75]{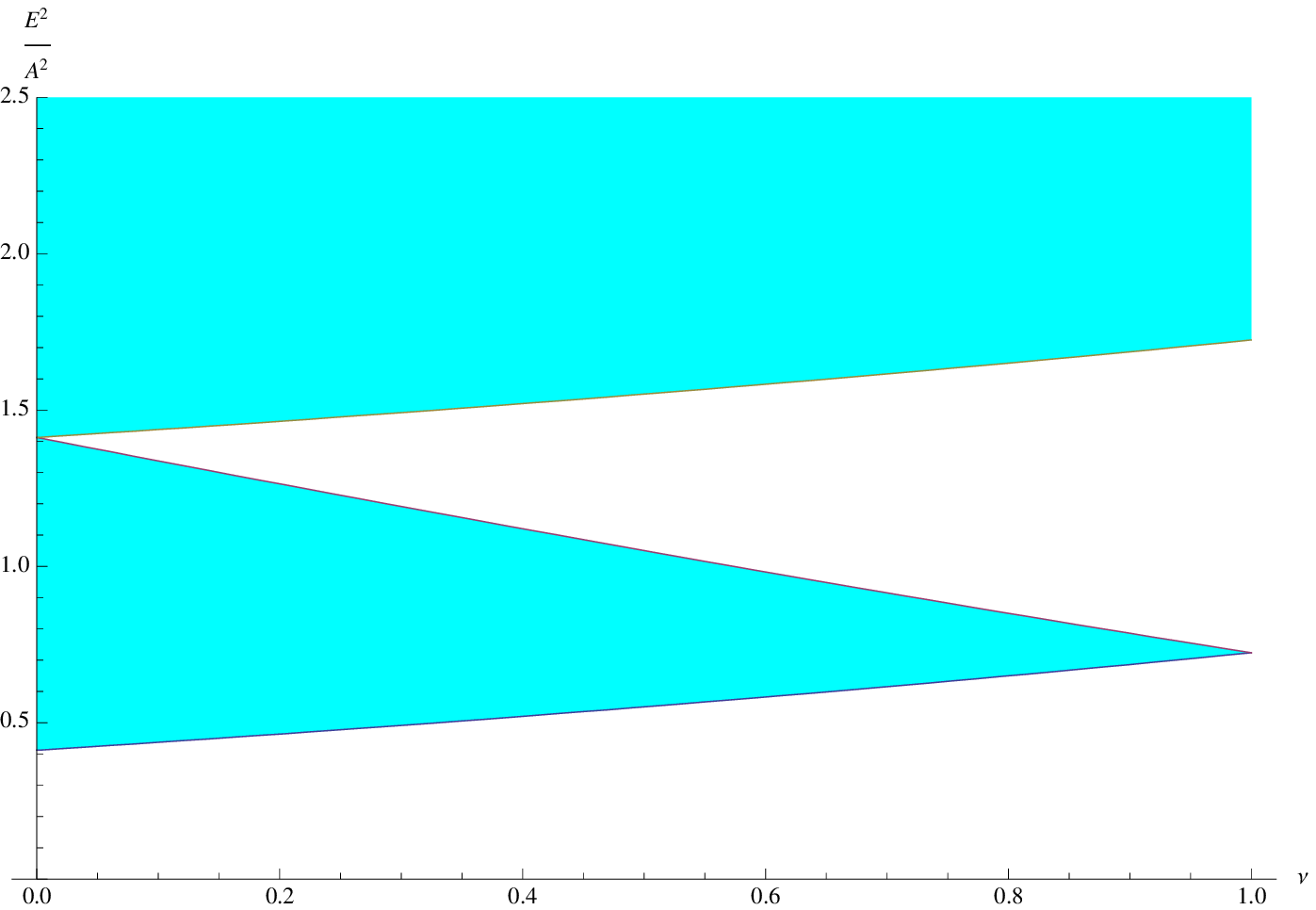}}
\caption{The spectrum of the Schr\"odinger operators $H_\pm$ with $\sigma(x)$
  given by the periodic condensate \protect\Eqref{kk-crystal}, as a function
  of  the elliptic parameter $\nu$. The spectrum has a single
  bound band, with band edges $A^2 {\rm cs}^2(b; \nu)$ and  $A^2 {\rm ds}^2(b;
  \nu)$, and a continuum band starting at  $A^2 {\rm ns}^2(b; \nu)$. When
  $\nu\to 1$, the bound band contracts to a single bound level at
  $A^2/\sinh^2b$, with the continuum edge at $A^2 \coth^2 b$. In the other
  limit, as $\nu\to 0$, the spectrum is pure continuum, with lower edge at
  $A^2 \cot^2 b$. This plot is for  $b=1$. These features can be
  clearly seen in the associated resolvents in the text.}
  \label{fig:kk-crystal-spectrum}
}

The resolvents for the potentials in (\ref{potentials-kk-crystal}) are derived
in Appendix~\ref{app:oldA}, and
the corresponding propagators are obtained by an inverse Laplace transform:
\bea
K_\pm(x; T)=\alpha(T)+\beta(T)  \left\{
\begin{matrix}
{\rm cn}^2(A\, x+b; \nu) \\
{\rm cn}^2(A\, x; \nu)
\end{matrix}
\right\}
\quad .
\label{periodic-propagator}
\eea
Note that the coefficients $\alpha(T)$ and $\beta(T)$ are the same for the two
propagators -- the only difference between $K_+$ and $K_-$ is the shift of the
$x$ dependence in the ${\rm cn}^2$ functions. The coefficients are most usefully
expressed in the convolution form:
\bea
\beta(T)&=& \frac{\nu\, A^2}{2}\, e^{-A^2 {\rm ds}^2(b; \nu)\, T} \int_0^Tdu\, \frac{e^{-A^2(\nu-1/2)u}}{\sqrt{\pi(T-u)}}\, I_0\left(\frac{A^2}{2}u\right)\\
\alpha(T)&=&\frac{1}{\nu}\left({\rm ds}^2(b; \nu) \, \beta(T)+\frac{1}{A^2}\, \frac{d\beta}{dT}\right) \quad .
\label{alphabeta}
\eea
When $\nu=1$ we recover the propagators for the kink-antikink pair
configuration in (\ref{kkbar-propagator}), while when $\nu=0$ we find
propagators for a homogeneous condensate of mass $A \,{\rm cot}(b)$.

An interesting special value of $\nu$ is the lemniscate case,
$\nu=\frac{1}{2}$, for which $\beta(T)$ takes the simple Bessel function form
\bea
\beta(T)=\frac{1}{4} e^{-A^2 {\rm ds}^2(b; \frac{1}{2})\, T}  \sqrt{\frac{\pi A^2 T}{2}}\, I_{\frac{1}{4}}\left(\frac{A^2 T}{4}\right)\, I_{-\frac{1}{4}}\left(\frac{A^2 T}{4}\right)\quad .
\label{n=1/2}
\eea
For a comparison with worldline numerics, we choose typical potential
parameters as they occur in the phase diagram \colH{of the GN model} at finite densities and
temperature: $b=1$, $\nu=0.9$, and $A=m/\coth(1)$. Since this typical form of
the potential is free of (near-)zero modes, it represents a direct test of the
quantitative reliability of worldline Monte Carlo for fermionic fluctuations
in generic backgrounds.  Figure \ref{fig:pathint} shows the expectation value
$\langle \exp(- \int V_-)\rangle$ versus the propertime in a minimum of the
potential $V_-$. Due to the absence of a zero mode, the values decrease
exponentially already for small propertimes, in contrast to the corresponding
single kink result in Figure \ref{fig:overlap}, and we see in Figure \ref{fig:pathint} that the
numerical and analytic results agree
very well. Even a small loop ensemble, of just $2000$ loops, yields tiny errors. By contrast, the first-
and second-order heat-kernel expansion, which is a standard analytic expansion
technique, fails badly for $T\gtrsim 1$.
For further quantitative comparison, we integrate the
resulting Lagrangian $\mathcal{L}_-$ over one period along the $x$ axis:
\begin{equation}
	\int_0^{2\mathrm K(0.9)/A} dx\ \mathcal L_-
	= -\frac{A}{4\pi}
		\begin{cases}
		6.81704\qquad\qquad ({\rm analytic})\cr
		6.818 \pm 0.001 \qquad ({\rm worldline\,\, numerics})
		\end{cases}
\end{equation}
where the upper value in braces is the analytic result, the lower one a
worldline-numerical estimate with 20 000 CM loops of 2000 ppl each. (The
result obtained for the integrated $\mathcal{L}_+$ is identical.)  In units of
$m$, the spatial average of $\mathcal L_-$ then is
\begin{eqnarray}
	\overline{\mathcal L_-}
	&:=& \frac{\kappa}{2 \mathrm K(0.9)}
	\int_0^{2\mathrm K(0.9)/\kappa} dx\ \mathcal L_-
	\nonumber\\
	&=& -\frac{m^2}{4\pi}
	\begin{cases}
	0.766857\qquad\qquad ({\rm analytic})\cr
	0.7669 \pm 0.0001  \qquad ({\rm worldline\,\, numerics})
	\end{cases}
\end{eqnarray}
Note that this is well above the corresponding value of the constant-field
solution, $-1/(4\pi)$, indicating that the constant field is the preferred
vacuum solution.  We conclude that a precision on the permille level or below
is straightforwardly achievable with worldline Monte Carlo algorithms for
fermionic systems in absence of (near-)zero modes.
\FIGURE[t]{
	\centering
	\includegraphics[width=0.66\columnwidth]{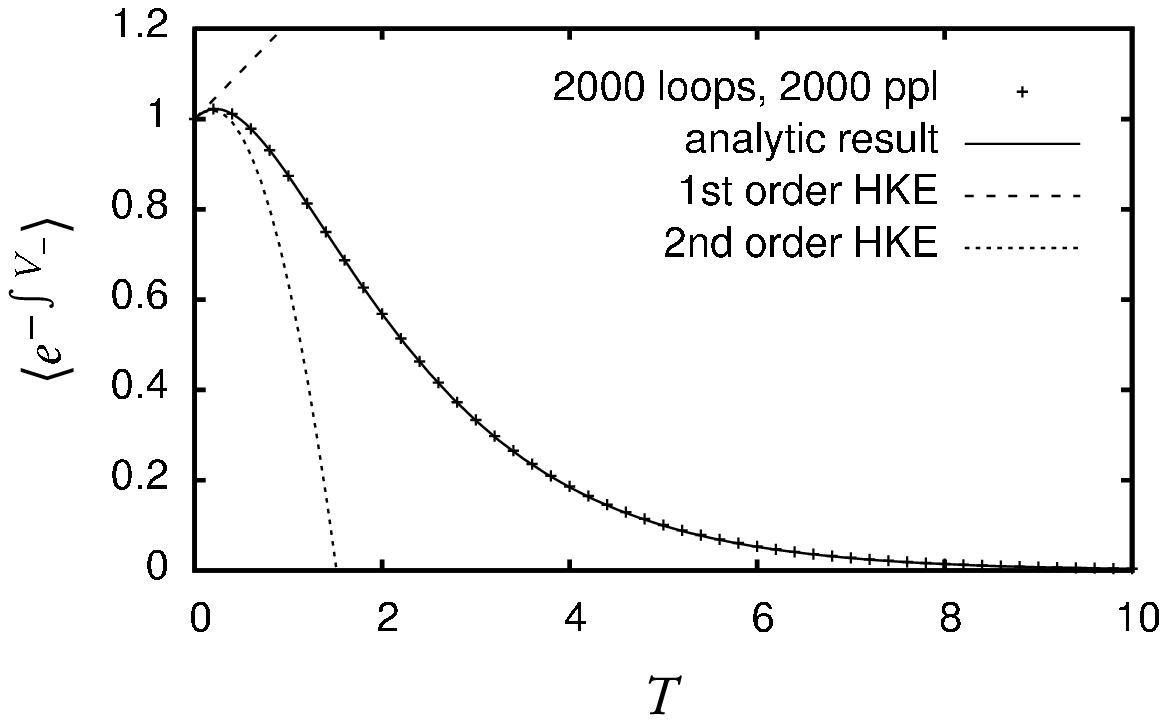}
	\caption{The worldline expectation-value versus the propertime in a
          minimum of the potential $V_-$ for $b=1$, $\nu=0.9$, and
          $A=m/\coth(1)$. The statistical error is too tiny to be plotted. Due
          to the absence of a zero mode, the values decrease exponentially very
          early, in contrast to the corresponding values for a single kink, as shown in Fig.
          \protect\ref{fig:overlap}. The figure above shows that the analytic heat-kernel
          expansion [the dashed and dotted lines] is not reliable for $T\gtrsim 1$. }
	\label{fig:pathint}
}

With the same loop ensemble we study the slightly more challenging
configuration $b = 2$ , $\nu = 0.9$ and $m = A\coth(2)$, which has
deeper minima, and obtain
\begin{equation}
	\int_0^{2\mathrm K(0.9)} dx\ \mathcal L_-
	= -\frac{A}{4\pi}
		\begin{cases}
		0.499945\qquad\qquad {\rm (analytic)}\cr
		0.49 \pm 0.03 \qquad {\rm (worldline\,\, numerics)}
		\end{cases}
\end{equation}
with the spatial average (in units of $m$)
\begin{equation}
	\overline{\mathcal L_-}
	= -\frac{m^2}{4\pi}
	\begin{cases}
	0.0901099 \qquad\qquad ({\rm analytic})\cr
	0.088 \pm 0.005 \qquad ({\rm worldline\,\, numerics})
	\end{cases}
\end{equation}
The statistical error is larger, as one might have expected, but can be
improved using a larger worldline ensemble. The agreement with the
analytic result is still very satisfactory.

\subsection{Kink crystal condensate}

An important special case corresponds to choosing the kink-antkink separation
to coincide with half the period of the crystal, in which case the antikinks
neighboring a given kink lie at the edges of the period. This corresponds to
choosing $b={\rm K}(\nu)$. Then the condensate in \Eqref{kk-crystal}
simplifies to
\bea
\sigma(x)=A\, \nu\, \frac{{\rm sn}(A\, x; \nu) {\rm cn}(A\, x; \nu)}{{\rm
    dn}(A\, x; \nu)}.
\label{k-crystal}
\eea
The associated Schr\"odinger potentials are
\bea
V_\pm=A^2\left( \nu -2\nu  \left\{
\begin{matrix}
(1-\nu)\,{\rm sd}^2(A\, x; \nu) \\
{\rm cn}^2(A\, x; \nu)
\end{matrix}
\right\}
\right) \quad .
\label{potentials-chiral}
\eea
The condensate and associated potentials are plotted in Figure \ref{fig:k-crystal}.
\FIGURE{
\centerline{\includegraphics[scale=0.5]{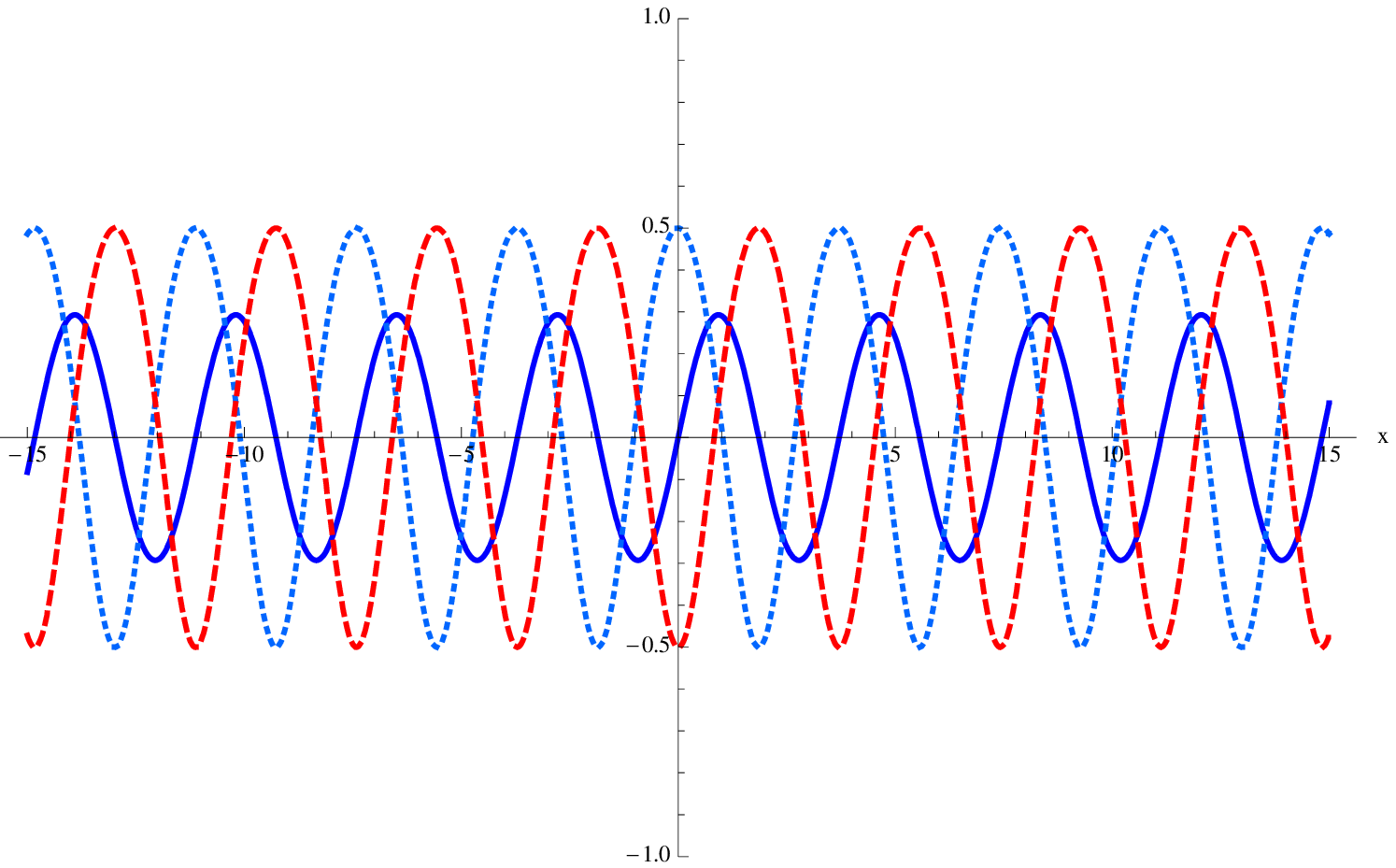}
\includegraphics[scale=0.5]{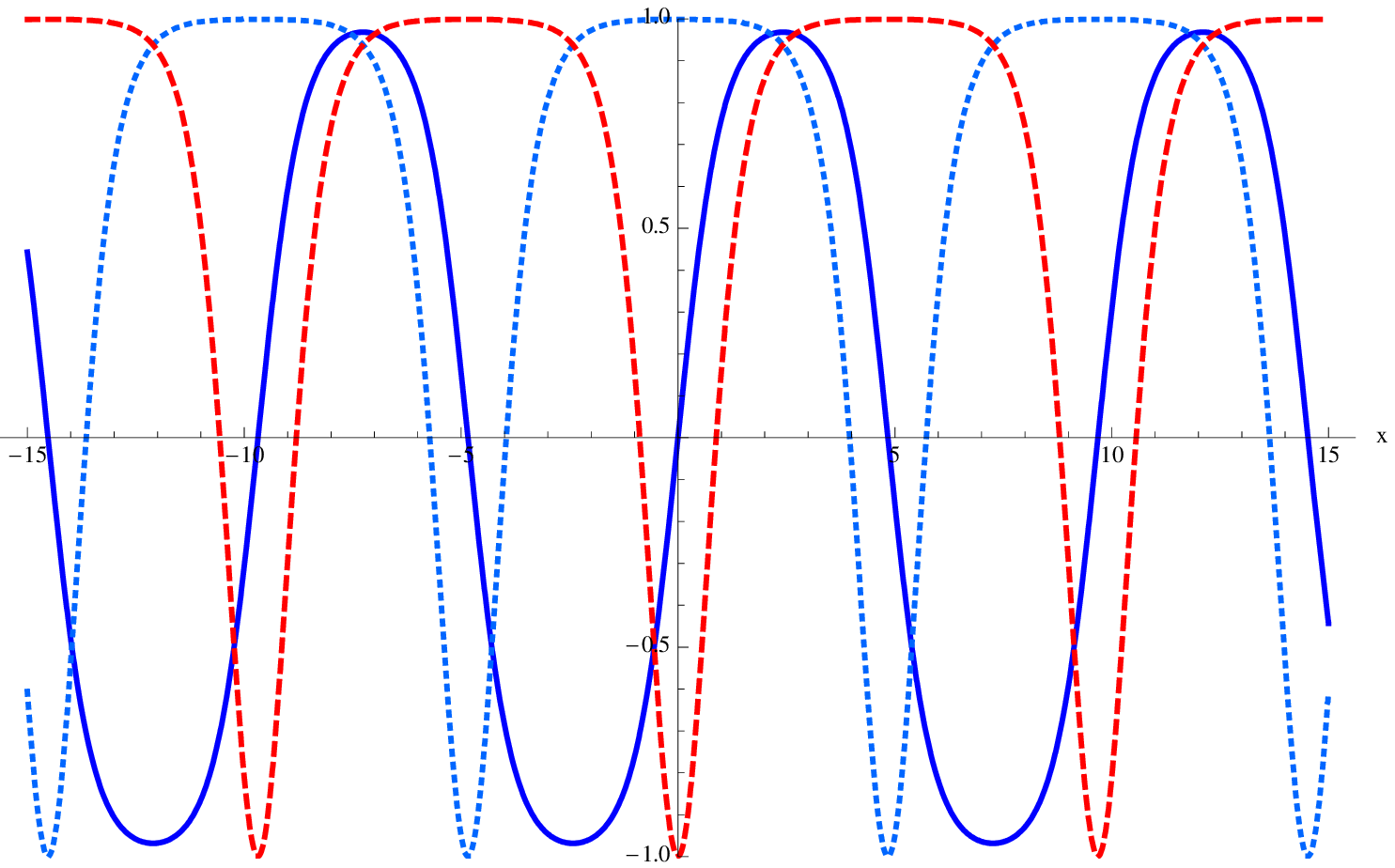}}
\caption{Plots of the periodic condensate $\sigma(x)$ [solid, dark blue, lines] in
  \protect\Eqref{k-crystal}, and the periodic potentials $V_\pm(x)$ in
  \protect\Eqref{potentials-chiral}, with elliptic parameter $\nu=0.5$ [first
  plot] and $\nu=0.999$ [second plot]. Both plots are for $A=1$. Note that the
  antikinks lie at the edge of the period, so that $V_-(x)$ [dashed, red, line] is localized on
  the kinks at the center of a period, while $V_+(x)$ [dotted, light blue, line] is localized on the
  antikinks at the edge of a given period.}
\label{fig:k-crystal}
}
\FIGURE{
\centerline{\includegraphics[scale=0.75]{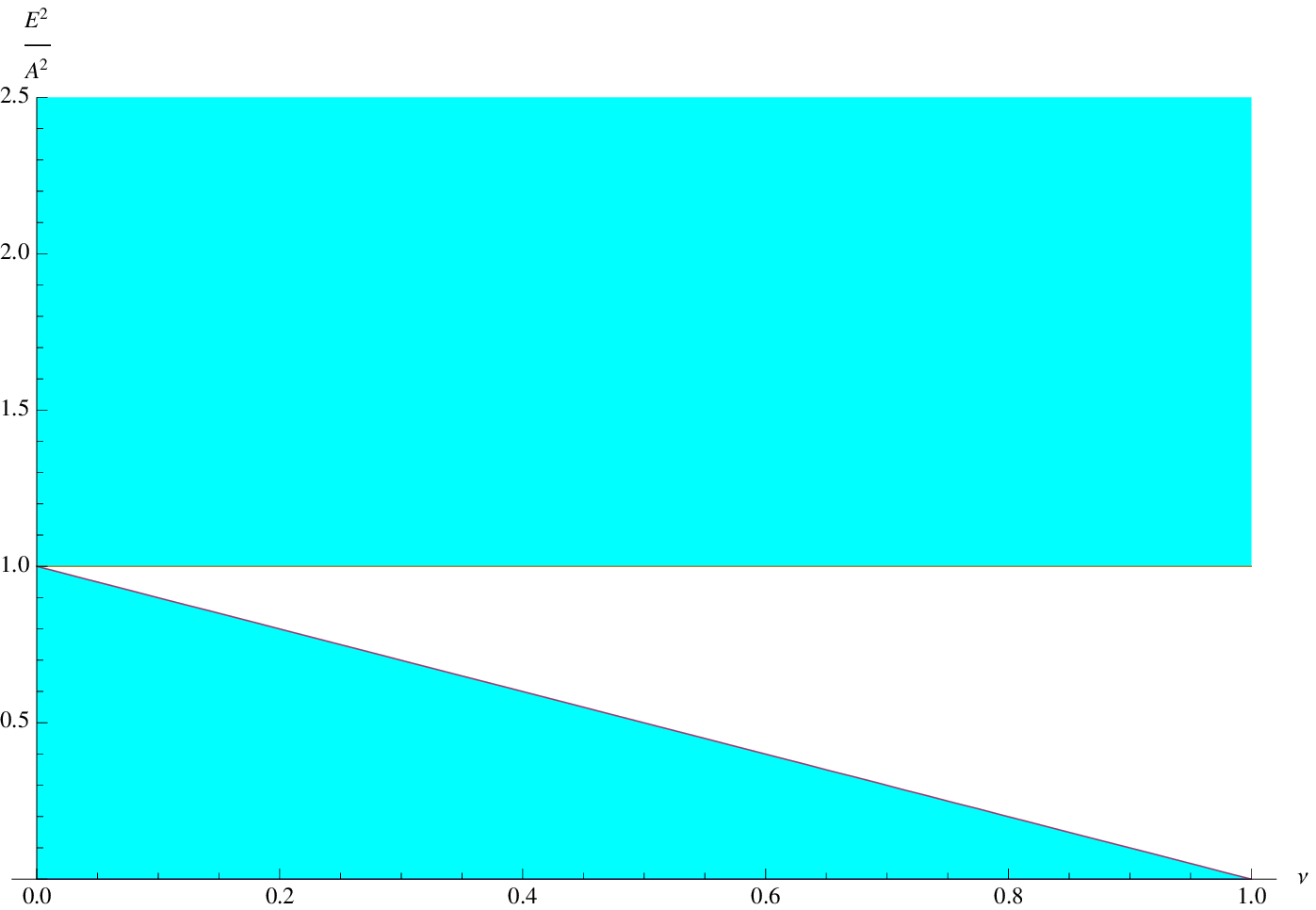}}
\caption{The spectrum of the Schr\"odinger operators $H_\pm$ with $\sigma(x)$
  given by the periodic condensate \protect\Eqref{k-crystal}, as a function of
  the elliptic parameter $\nu$. The spectrum has a single bound band, with
  band edges $0$ and $A^2 (1-\nu)$, and a continuum band starting at
  $A^2$. When $\nu\to 1$, the bound band contracts to a single bound level at
  $0$. In the other limit, as $\nu\to 0$, the spectrum is pure continuum, with
  lower edge at $0$.}
  \label{fig:k-crystal-spectrum}
}

This system has a band spectrum with band edges at $0$, $A^2(1-\nu)$, and
$A^2$, as shown in Figure \ref{fig:k-crystal-spectrum}. Note that the bottom
of the lowest band is at 0, for all $\nu$, and that there is just one bound
band. The corresponding heat kernels $K_\pm(x; T)$ can be obtained from those in the previous
section by the simple substitution $b\to {\rm K}(\nu)$.

%\newpage

\subsection{Worldline ensembles from Hybrid Monte Carlo}
\label{sec:HMC}

Since the standard worldline ensembles generated with respect to the
  free worldline action can lead to an overlap problem for the special case of
  a (near-)zero mode in the fermion spectrum, let us explore a different
  route: we generate the ensemble with respect to the {\it full action} using a
  Hybrid Monte Carlo (HMC) algorithm. In principle, worldline generation with
  the full action would also be possible with a local update algorithm such as
  the Metropolis update. However, due to the fact that the worldlines are
  one-dimensional, local updates suffer from a long auto-correlation time, as
  observed in \cite{Gies:2003cv}. This is avoided by the global HMC update.

The quantity which will be central for the HMC simulation is given by
\begin{eqnarray}
W(x,T) &=&  - \, \frac{d}{dT} \, \ln \, K_\pm (x,T) \; = \;
\frac{ \langle x \vert (-\partial_1^2 + V_\pm) e^{-T (-\partial_1^2 + V_\pm)}
\vert x \rangle }{ \langle x \vert e^{-T (-\partial_1^2 + V_\pm) }
\vert x \rangle } \; ,
\label{eq:w10} \\
 K_\pm (x,T) &=&  \langle x \vert e^{-T (-\partial_1^2 + V_\pm) }
\vert x \rangle \; .
\end{eqnarray}
This quantity is given  by means of an expectation value
with respect to the full loop ensemble:
\begin{equation}
W(x,T) \; = \; \Bigl\langle -\partial_1^2 + V_\pm\Bigr\rangle _{xF} \; ,
\label{eq:w11}
\end{equation}
where the loops are generated with the probability
\begin{equation}
P_{\text{F}}[x]= \delta\left( x_{\text{CM}} - \int_0^T d\tau x(\tau) \right) \,
 \exp \Bigl[-\frac{1}{4} \int_0^T   d\tau \, \dot{x}^2(\tau) - \int _0^T
d\tau \,   V_\pm(x(\tau))  \Bigr] .
\label{eq:w12}
\end{equation}
The strategy is to generate statistically important loop configurations $\{
x(\tau) \}$ using HMC methods and to estimate $W(T,x)$ in (\ref{eq:w11}).  The
advantage of this approach is that the loop \colH{ensemble already incorporates} features
of the potentials $V_\pm$, and that far fewer configurations are required for
a reliable estimate of expectation values than for free loop clouds. The
disadvantage is that more computational resources are necessary to
generate the HMC loop ensembles.  Ultimately, it will depend on the particular potential
whether the additional numerical effort is justified. We will argue that this
is the case for a Hamiltonian $H_\pm=(-\partial_1^2 + V_\pm) $ which possesses
zero modes.

Once a numerical estimate for $W(x,T)$ is obtained, the heat kernel $K_\pm$ can
be reconstructed by integration of (\ref{eq:w10}) with respect to the
propertime.  Note that the kernel satisfies the boundary condition $ K_\pm
(x,0) = 1$. We therefore find:
\begin{equation}
K_\pm (x,T) \; = \; \exp \left\{ - \; \int _0^T d\tau \; W(x,\tau)
\; \right\} \; .
\label{eq:w15}
\end{equation}
Let us briefly comment on the implementation of the discretized loops.
As before, the propertime interval $\tau = [0,T]$ is represented using
$N$ points:
\begin{equation}
\tau \; = \; n \, d\tau \; , \hbox to 1cm {\hfill }
d\tau \; = \; \frac{T}{N} \; , \hbox to 1cm {\hfill }
n = 1 \ldots N \; , \hbox to 1cm {\hfill } N \; \hbox{odd} \; .
\label{eq:w16}
\end{equation}
For a simplification of the notation, we will assume that $N$ is odd.
We base the HMC algorithm on a discrete Fourier representation of the
  worldlines, which can easily account for the worldline periodicity and leads
  to uncoupled degrees of freedom for $V_\pm=0$. Hence, considering the
Fourier coefficients as degrees of freedom implies that loop ensembles can be
generated with little autocorrelation as long as the influence of the
potential is not too significant.\footnote{An HMC algorithm could
    equally be based on the {\em v loop} algorithm \cite{Gies:2003cv}, since
    it features the same necessary properties. By contrast, an HMC
    generalization of the {\em d loop} algorithm \cite{Gies:2005sb} which is
    based on an iterative loop construction seems not to be straightforward.} For CM
loops, we have the representation:
\begin{eqnarray}
x(\tau ) \; =  x_\cm+\frac{\sqrt{T}}{N} \;
\sum _{k=-\frac{N-1}{2}}^{\frac{N-1}{2}} c_k \; \exp \left( i \, \frac{2 \pi }{T} \, k \,
  \tau \right) \; ,
\hbox to 1cm {\hfill } c_{-k} = c_k^\ast,\quad c_0=0.
\label{eq:w17}
\end{eqnarray}
In the case of CP loops, for which $x_\cp$ is the common point, we
analogously find
\begin{equation}
x(\tau ) \; = \; x_\cp \; + \; \frac{\sqrt{T}}{N} \;
\sum _{k=-\frac{N-1}{2}}^{\frac{N-1}{2}} c_k \; \left[ \exp \left( i \,
\frac{2 \pi }{T} \, k \, \tau \right) \, - 1 \, \right]
\; .
\label{eq:w18}
\end{equation}
With these representations, the worldline kinetic term is Gau\ss ian
in the weights $c_k$:
\begin{eqnarray}
\frac{1}{4} \int _0^T d\tau \; \dot{x}^2(\tau) &\to&
\frac{1}{4} \sum_{n=1}^{N} \frac{( x(\tau+d\tau) - x(\tau))^2 }{d\tau }
\; = \;
 \, \sum_{k} c_k \, c_{-k} \, \sin^2 \left( \frac{\pi}{N} \, k \right).
\label{eq:w19}
\nonumber
\end{eqnarray}
The task is now to generate statistical ensembles of the degrees of
freedom
$$
c_k \; , \hbox to 1cm {\hfill } k \; = \; - \frac{N-1}{2},  \ldots, -1, \;
1, \ldots  \frac{N-1}{2} ,
$$
which are distributed according to, e.g.,  the $\cm$ probability distribution
(\ref{eq:w12}). Note that the $\delta $-function constraint
has been exactly incorporated leaving us with $2 \, \frac{N-1}{2} = N -1$
degrees of freedom, parameterized by the coefficients $c_k$.
For the HMC approach, it is highly advisable to introduce real valued
degrees of freedom. We therefore define the coefficients $d_k \in \mathbb{R}$
by
\begin{equation}
d_k \; = \; c_k + c_{-k} \; , \hbox to 1cm {\hfill }
d_{-k} \; = \; i\, (c_k - c_{-k}) \; , \hbox to 1cm {\hfill }
k=1 \ldots \frac{N-1}{2} \; .
\label{eq:w19b}
\end{equation}
With this definition the loop coordinates, for instance, for the CM
loops (\ref{eq:w17}) are manifestly real valued:
\begin{equation}
x(\tau ) \; = \; x_\cm \; + \; \frac{\sqrt{T}}{N} \;
\sum _{k=1} ^{\frac{N-1}{2}} \left\{ d_k \;\cos \left(
\frac{2 \pi }{T} \, k \, \tau \right)
\; + \; d_{-k} \sin \left( \frac{2 \pi }{T} \, k \, \tau \right)
\right\} \; .
\label{eq:w19c}
\end{equation}
The next step is to introduce the canonical momentum $\pi _k\in \mathbb{R}$
for each degree of freedom $d_k$. The corresponding HMC Hamiltonian is
defined by
\begin{eqnarray}
{\cal H} (\pi_k, d_k) &=& \frac{1}{2} \sum _k \left[ \pi^2_k  \; + \;
{\cal V}(d_k) \, \right],
\label{eq:w20} \\
{\cal V}(d_k) &=& \frac{1}{4} \int_0^T d\tau \, \dot{x}^2(\tau)
+ \int _0^T d\tau \,   V_\pm(x(\tau)) \; .
\label{eq:w21}
\end{eqnarray}
The degrees of freedom experience an evolution with the so-called
HMC time $u$. This time evolution is determined by the Hamilton equations
of motion:
\begin{equation}
\frac{d}{du} \, d_k \; = \; \pi _{k} \; , \hbox to 1cm {\hfill }
\frac{d}{du} \, \pi_k \; = \; - \;
\frac{ \partial {\cal V} (d_k) }{ \partial d_{k} } \; .
\label{eq:w22}
\end{equation}
The important property of any solution $d_k(u)$, $\pi _k(u)$ is that
the HMC energy is conserved:
$$
\frac{d}{du} \; {\cal H} \Bigl(\pi_k(u), d_k(u)\Bigr) \; = \; 0 \; .
$$
The initial momenta $\pi_k(u=0)$ are selected at random from
a Gaussian distribution. The initial coordinates $c_k(u=0)$ are
given by their actual ensemble values. With these initial conditions,
the values at $u=u_f$ are {\it approximately } obtained with the help
of a Leapfrog integration of the equations of motion (\ref{eq:w22}).
Because the equations of motion are not solved exactly, the
HMC energy ${\cal H}$ is not exactly conserved. The idea central
to the HMC approach is to accept the values $d_k(u_f)$ as  new
members of the Markov chain with the probability
\begin{eqnarray}
p =  \hbox{min} \, \Bigl(1 , \, \exp \{ - \, \Delta H \} \, \Bigr)  ,
%\label{eq:w23} \\
\hbox to 1cm {\hfill }
\Delta H \, = \, {\cal  H} \Bigl[ \pi_k(u_f),d_k(u_f) \Bigr]
 - {\cal  H} \Bigl[ \pi_k(0),d_k(0) \Bigr] \, .
\label{eq:w24}
\end{eqnarray}
Since the equations of motion are satisfied to a good extent due to
the Leapfrog integration, the value $c_k(u_f)$ can
be largely different from $c_k(0)$ (implying good ergodicity)
while the acceptance rate is still high.

\FIGURE{
\centerline{
\includegraphics[scale=0.4]{energ_v}
\includegraphics[scale=0.4]{wtx2}}
\caption{Left panel: the HMC energy as a function of step size $du$ of the
Leapfrog integration. Right panel: the function $W(x=0,T)$  for the kink
configuration obtained with HMC loops. The large-$T$ behavior is
dominated by the ground-state energy, cf. \protect\Eqref{eq:w25}. Sufficiently
many points per loop $N$ are required in order to observe the approach
to the zero-energy mode of the kink.}
  \label{fig:hmc}
}
In order to test the HMC approach, we revisit the kink
background field (\ref{kink}) (with $A=1$ in the following).
The HMC final time is given by $u_f = N_\mathrm{tra} \, du $ where
$N_\mathrm{tra}$ is the number of steps of the Leapfrog integration.
The step size $du$ is chosen to obtain an acceptance rate between
20\% and 80\%. Figure~\ref{fig:hmc} (left panel) shows the
HMC energy as function of the step size $du$ for several values
for the proper time. We find that a value $du \approx 0.3$
provides a good compromise between ergodicity and acceptance rate.
Using the HMC ensembles for CM loops, the function $W(x,T)$ is
estimated and the heat kernel is reconstructed with the help
of the proper time integration (see \Eqref{eq:w15}).
Note that because of (\ref{eq:w10}), we find for large values of $T$
\begin{equation}
W(x,T) \; \approx \; E_0 \, + \, (E_1-E_0) \,
\frac{ \vert \langle x \vert 1 \rangle
\vert ^2 }{\vert  \langle x \vert 0 \rangle \vert ^2 }
\, \mathrm{e}^{ - (E_1- E_0) \, T }\quad
\hbox{($T$ large)},
\label{eq:w25}
\end{equation}
where $E_0$, $E_1$ are the energies of ground and first excited states,
and $\vert 0 \rangle $ and $\vert 1 \rangle $
are the corresponding wave functions.
In theories with a gap between the first excited and the ground state,
the latter equation implies that $W(x,T)$
rapidly approaches a constant, i.e., the ground state energy $E_0$,
for large $T$. In the worst case
scenario of a zero mode, $W(x,T)$ approaches zero and a positive constant
otherwise.
Figure~\ref{fig:hmc} (right panel) shows the result for $W(x=0,T)$ for
the kink configuration. Indeed, the function $W(x,T)$ rapidly approaches
a constant for large $T$. The overlap problem, discussed in the previous
subsections, has been solved by the use of HMC loop ensembles.
Note, however, that, also in the present case, the number of points per loop
$N$ must be large enough: for coarse loops, the numerical estimate
approaches a negative constant thus erroneously suggesting a negative ground state
energy and an instability of the operator. A sufficiently fine
representation or an improved integration over loop coordinates
might resolve this problem.

\FIGURE{
\centerline{
\includegraphics[scale=0.58]{kink_heatkernel_CMloops3}
}
\caption{HMC results for the heat kernel for the kink configuration in
  comparison with the result obtained from free loop ensembles (all used
  ensembles are $\cm$ loops). Significantly fewer HMC loops are required for a
  reliable estimate of the large-propertime behavior. }
\label{fig:hmc_comp}
}
Finally, we present our final result for the reconstructed heat kernel
in Fig.~\ref{fig:hmc_comp}. We obtain very good results with HMC
using only a modest number of points per loop.

To conclude this subsection, the overlap problem of
the free loop cloud method is indeed solved by adopting the HMC
approach. Note, however, that a precise estimate of the effective action
in the case of a (near-)zero mode is still a numerical challenge.

\section{Worldline approach at finite temperature and chemical potential}

Even though introducing finite temperature and chemical potential appears
straightforward in the worldline approach, special attention has to be paid to
combining this straightforward formalism with Monte Carlo algorithms. The
reason is that an analysis of field theories at finite temperatures and
densities often requires knowledge about the analytic structure of propagators
etc. in the complex energy-momentum plane, or similar conjugate variables such
as the propertime. By contrast, Monte Carlo algorithms require a positive
real action for importance sampling, and this restricts the computable information
to lie on the real axis of given variables.

Since this conflict of interests between thermal field theory at finite
density and Monte Carlo algorithms is general and not only restricted to the
Gross-Neveu model, we discuss this problem in more detail in the following.

\subsection{Temperature in the worldline representation}

Consider the worldline formulation of a general quantum-field theoretic
amplitude. Finite temperature can easily be implemented with the aid of the
Matsubara formalism: The Euclidean time, say along the $D$th direction, is
compactified to the interval $[0,\beta]$ with periodic boundary conditions for
worldline fluctuations for bosonic fields and antiperiodic boundary conditions
for fermionic degrees of freedom. Here $\beta=1/\mathcal{T}$ is the inverse
temperature. (Note that we reserve the symbol $T$ for propertime,
  whereas the temperature is denoted by $\mathcal T$.) As a consequence, the
worldlines can also wind around the time dimension. It is convenient to write
a given loop $x(\tau)$ with winding number $n$ as sum of a loop with no
winding, $\tilde x(\tau)$, and a translation in time running from zero to $n
\beta$ with constant speed,
\begin{equation}
	x_\mu(\tau)
	= \tilde x_\mu(\tau) + n\beta\frac{\tau}{T} \delta_{\mu D}.
	\label{eq:winding}
\end{equation}
The path integral over the different winding number sectors labeled by
$n$ factorizes for static configurations, yielding
\begin{equation}
	\int_{x(0)=x(T)}\mathcal Dx\ e^{-\int_0^{T}d\tau
          \frac{\dot x^2}{4}} \cdots\
	= \sum_{n=-\infty}^\infty (-1)^n e^{- \frac{n^2\beta^2}{4T}}
		\int_{\tilde x(0)=\tilde x(T)}\mathcal D\tilde x
			\ e^{-\int_0^{T}d\tau \frac{\dot{\tilde x}^2}{4} }
			\ \cdots. \label{eq:windingsum}
\end{equation}
where the factor $(-1)^n$ implements antiperiodic boundary conditions for the
present fermi\-on fluctuations; it would be absent for bosons here and in the
following.  A finite-tempera\-ture generalization of a vacuum
propertime expression is straightforwardly obtained from a simple insertion
into the propertime integral,
\begin{equation}
  \int_0^\infty dT\, \dots \to \int_0^\infty dT\, \left( \sum_{-\infty}^\infty
    (-1)^n e^{- \frac{n^2\beta^2}{4T}} \right) \dots
\end{equation}
where the ellipsis represent, for instance, a heat-kernel expression or a
worldline expectation value etc. The winding number sum over $n$ is directly
related to the Matsubara sum by a Poisson transformation. Quantum and thermal
fluctuations are separated by the different winding number sectors:
the $n=0$ term is identical to the vacuum expression, while the $|n|\geq 1$ sectors are
purely thermal contributions which vanish in the vacuum limit $\beta\to \infty$.

Let us stress that the present thermal formalism not only holds in general but
is also completely robust when combined with worldline Monte Carlo techniques,
\colH{as has been used in the context of the Casimir effect at finite
  temperature \cite{Gies:2008zz}.}. The winding number sum converges rapidly
in the relevant low- and intermediate temperature range; at high temperatures
where increasingly more winding sectors would have to be resummed, a Poisson
resummation to Matsubara sums can be performed, such that rapid convergence is
guaranteed for all temperatures. In particular, analytical or numerical
knowledge of the heat kernel on the real $T$ axis is sufficient for a reliable
estimate of thermodynamic amplitudes.  {For example, it is straightforward} to
derive the critical temperature for symmetry restoration in the Gross-Neveu
model within the present formalism, yielding the well-known result
\begin{equation}
	\beta_\mathrm c = \pi e^{-C} m^{-1} \approx 1.76387 m^{-1}, \quad
	{\mathcal T}_\mathrm c = 1 / \beta_\mathrm c \approx 0.566932 m \label{eq:Tc}
\end{equation}
in units of $m$ \cite{Wolff:1985av}. For temperatures larger than the critical
temperature ${\mathcal T}_\mathrm c$, the potential $\sigma$ vanishes and the
spontaneously broken discrete chiral symmetry is restored.

\subsection{Chemical potential in the worldline representation}

Introducing a finite chemical potential via the grand-canonical partition
function leads us to the microscopic action of the Gross-Neveu model at finite
density,
\begin{align}
	S
	&=
	\int d^Dx \left(
	-\bar\psi
	\left( \fss\partial + \mu \gamma_D \right) \psi
	+ \frac{g^2}{2\Nf} ( \bar\psi \psi )^2
	\right),
\end{align}
in $D$ dimensions. The resulting contribution to the effective action from
fermionic fluctuations reads analogously to \Eqref{eq:DelG0}
\begin{equation}
  \Delta\Gamma[\sigma]=-\frac{\Nf}{2} \Tr \ln (-\vec\partial^2 - (\partial_D
  +\mu)^2 +  \sigma^2  - i \fss{\partial} \sigma  ) \label{eq:DelG0b}.
\end{equation}
Since the chemical potential occurs in the form of the $D$th component of an
imaginary \colH{Euclidean} gauge potential in the effective action, it is
tempting to jump directly to the corresponding worldline representation in the
presence of gauge fields, which are coupled to the worldline in the form of a
Wegner-Wilson loop. In other words, the free worldline action receives a
further contribution of the form
\begin{equation}
e^{-\frac{1}{4} \int_0^T d\tau \dot{x}^2} \to
e^{-\frac{1}{4} \int_0^T d\tau \dot{x}^2 -\mu \int_0^T d\tau \dot{x}_D}
\to
(-1)^n e^{- \frac{n^2 \beta^2}{4 T}}e^{-\frac{1}{4} \int_0^T d\tau \dot{\vec
    x}^2 -\mu\beta n },
\end{equation}
where in the last step we have used the decomposition of worldlines into
different winding number sectors $n$ on the finite-temperature cylinder. Since
also negative winding numbers are allowed, it is already apparent that the
winding number sum may suffer from convergence problems at large
$\beta\mu$. This issue will be discussed in detail in the next subsection; let
us simply state here that convergence problems are absent for $\beta\mu<\pi$
which we will assume in the following.

Decomposing the winding number sum into the vacuum sector $n=0$, and the
temperature and density corrections $|n|\geq1 $, implying a decomposition of
the effective action (or grand potential) into a  vacuum part $\Gamma$, and a thermal
part $\Delta\Gamma_{\beta,\mu}$:
\begin{equation}
\Gamma_{\beta,\mu}=\Gamma+\Delta\Gamma_{\beta,\mu},
\label{eq:Gthermo0}
\end{equation}
with $\Gamma$ given by \Eqref{eq:Gren}, we end up with the worldline
representation for the thermal contribution at finite density
\begin{equation}
	\Delta\Gamma_{\beta,\mu}
	=
	\int d^Dx
	\frac{\Nf d_\gamma}{(4\pi)^{D/2}}
	\int_0^\infty\frac{dT}{T^{1+D/2}}
	\biggl(
	\sum_{n=1}^\infty (-1)^n  e^{-\frac{\beta^2 n^2}{4T}}
	\cosh(\mu \beta n)
	\biggr)
	\Bigl\langle
	e^{-\int_0^Td\tau \sigma^2} \Phi[\sigma] \Bigr\rangle.
	\label{eq:Gthermo}
\end{equation}
For $\beta\mu<\pi$, this representation is reasonably converging and
can numerically be implemented rather straightforwardly. Naively
interchanging summation and integration yields rapidly converging
propertime integrals for any given $n$; however, the $n$ sum, which is
reminiscent of a fugacity expansion, then diverges for
$\beta\mu>\pi$. Of course, \Eqref{eq:Gthermo} can immediately be used
for a finite-density expansion in $\beta\mu$ to any order. It is also
straightforwardly applicable to the case of imaginary chemical
potential where all convergence issues are absent.

As an application of \Eqref{eq:Gthermo}, let us briefly revisit the
phase diagram for a homogeneous $\sigma$ condensate. The worldline
expectation value for this is trivial, since $ \exp (-\int V_{\pm})
=\exp(-T \sigma^2)$, for any single worldline. This case
 serves as a simple but instructive test for the (non-worldline) numerics of
our algorithms, and for the representation \Eqref{eq:Gthermo}.  In
Fig.~\ref{fig:oldpd}, the phase diagram obtained by numerically
minimising the full action with respect to $\sigma$ for a given
chemical potential and temperature is shown.  In the accessible
domain, that is for $\beta \mu < \pi$, the diagram agrees with the
result of \cite{Wolff:1985av}, as it should.  At $\mu = 0$, we recover
the critical temperature in \Eqref{eq:Tc}.  Even close to the line
$\beta\mu = \pi$, i.e. temperature ${\mathcal T}=\mu/\pi$, below which the
propertime integral has convergence problems, we obtain a stable
result.
\FIGURE[t]{
\centering
\includegraphics[width=.96\columnwidth]{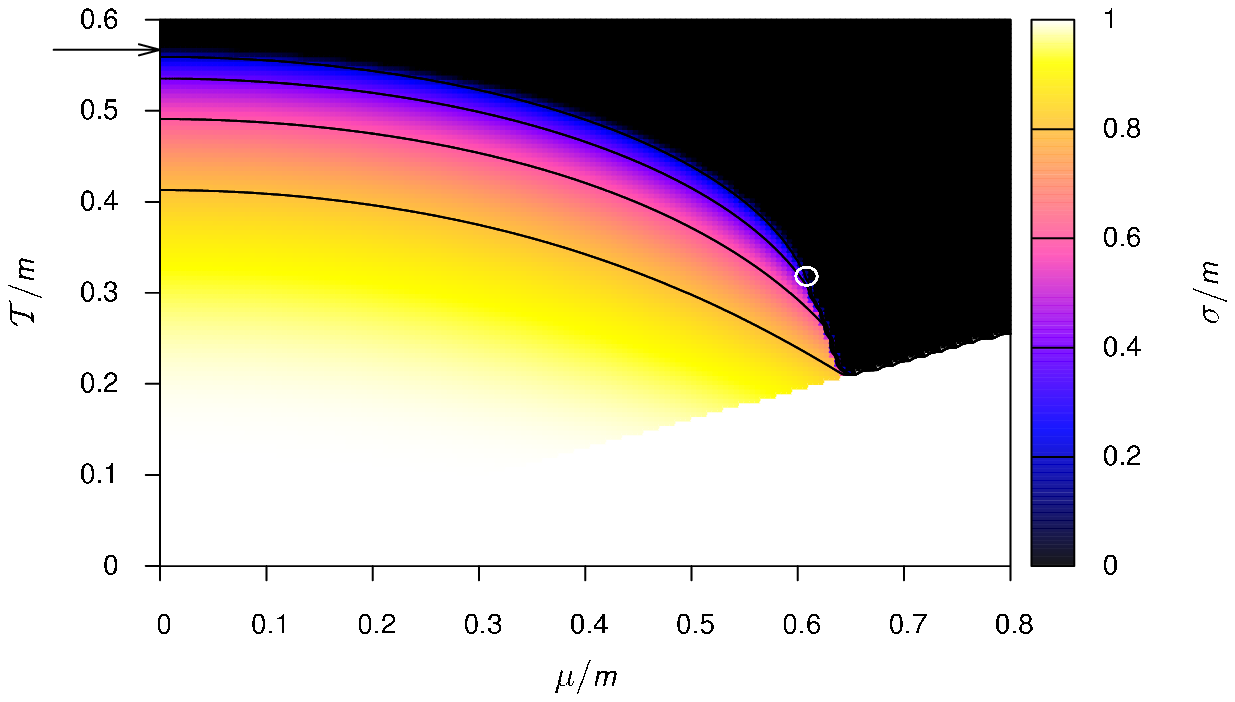}
\caption{Phase diagram for a constant potential
(cf. \protect\cite{Wolff:1985av}) computed by numerically minimising the full
action with respect to $\sigma$. The colour corresponds to the
resulting $\sigma$ value, the phase of restored chiral symmetry is in
black.  The tricritical point is marked by a white circle, located
between phase transitions of second order (the upper boundary of the
blue region) and transitions of first order (discrete jump of the
colour/$\sigma$ value).  The line of second order transitions extends
to the $\mu = 0$ axis, where it is marked by an arrow at the value
${\mathcal T}_\mathrm c \approx 0.567$ \protect\Eqref{eq:Tc}.  The blank region
below the ${\mathcal T} = \mu / \pi$ line is inaccessible to the given expression
for the action as discussed in the text.  }
\label{fig:oldpd}
}

\subsection{High densities and low temperatures }

Let us further investigate the instability for $\beta \mu >\pi $.
For this purpose, we first address the homogeneous background
$\sigma$=constant, $ \Phi[\sigma] =1$, and extend the findings to the general
case at the end of this subsection.

The (regularized) effective action is given by
\begin{equation}
	\Gamma_{\beta,\mu}
	= \frac{1}{2}
	\int d^Dx
	\frac{\Nf d_\gamma}{(4\pi)^{D/2}}
	\int_{1/\Lambda^2}^\infty\frac{dT}{T^{1+D/2}}
	\biggl(
	\sum_{n=-\infty}^\infty (-1)^n  e^{-\frac{\beta^2 n^2}{4T}}
	e^{ - \mu \beta n}
	\biggr) e^{-T \sigma ^2} .
	\label{eq:h1}
\end{equation}
In order to perform the sum over the windings $n$, we introduce
a Hubbard-Stratonovich transformation by
\begin{equation}
(-1)^n  e^{-\frac{\beta^2 n^2}{4T}} e^{ - \mu \beta n} \; = \;
\sqrt{ \frac{4\pi T }{ \beta ^2 } } \, \int d\nu \;
e^{i 2 \pi n \, \nu } \, e^{- T [\frac{2\pi}{\beta } (\nu +\frac{1}{2}) - i \mu ]^2 }.
\label{eq:h2}
\end{equation}
Inserting (\ref{eq:h2}) in (\ref{eq:h1}) yields for the $T$ integral
\begin{equation}
	\Gamma_{\beta,\mu}
	= \frac{1}{2}
	\int d^Dx
	\frac{\Nf d_\gamma}{(4\pi)^{D/2}}
\int_{1/\Lambda^2}^\infty\frac{dT}{T^{1+D/2}} \sqrt{ \frac{4\pi T }{ \beta ^2 } }
\sum_{n=-\infty}^\infty  \int d\nu \;
e^{i 2 \pi n \, \nu } \, e^{- T [\frac{2\pi}{\beta } (\nu +\frac{1}{2}) - i
  \mu ]^2 }
e^{-T \sigma^2} .
\label{eq:h3}
\end{equation}
{The IR problem for large values $T$ has become apparent: the $T$ integration
converges provided
\begin{equation}
\mathrm{Re} \, \left[ \left( \frac{\pi}{\beta } -i\mu \right)^2 + \sigma ^2
\right] > 0 .
\end{equation}
For imaginary chemical potential this condition is always satisfied. But for
real chemical potential we find the condition
\begin{equation}
\beta\mu<\sqrt{\pi^2+\beta^2\sigma^2}.
\label{betamu}
\end{equation}
Allowing for the possibility of a vanishing condensate, or in the more general
case for the existence of a zero mode in the spectrum, we restrict to chemical
potentials satisfying the bound
\begin{equation}
\beta \, \mu < \pi .
\label{eq:h4}
\end{equation}
}

The basic idea of the present subsection is to consider the case
(\ref{eq:h4}), then to derive an analytic expression for the effective action,
and to perform an analytic continuation to values $\mu > \pi/\beta $.
For a homogeneous background, we find that this approach recovers
the well-known result familiar from solid state physics.

Hence, let us confine ourselves to values $\mu $ satisfying
(\ref{eq:h4}), and perform the $T$ integration first.
For this purpose, we introduce the ``density of states'' $\rho _0 $ by
\begin{eqnarray}
\frac{1}{T^{(D-1)/2} } \, e^{-T \sigma ^2} &=& 2 (4\pi )^{(D-1)/2}
\int _0^\infty dE \, E \, \rho _0(E) \, e^{-TE^2} \, ,
\label{eq:h5} \\
\rho _0 (E) &=& \int \frac{ d^{D-1} k}{ (2\pi )^{D-1} } \;
\delta \Bigl( E^2 - \sigma ^2 - k^2 \Bigr) \, .
\label{eq:h6}
\end{eqnarray}
Rearranging (\ref{eq:h3}) with the help of (\ref{eq:h5}-\ref{eq:h6})
yields
\begin{equation}
	\Gamma_{\beta,\mu}
	= \Nf d_\gamma \int d^Dx \int_0^\infty dE \, E \, \rho _0(E) \,
\frac{1}{\beta } \sum_{n=-\infty}^\infty
\int_{1/\Lambda^2}^\infty\frac{dT}{T}
\int d\nu \;
e^{i 2 \pi n \, \nu } \, e^{- T [\frac{2\pi}{\beta } (\nu +\frac{1}{2}) - i
  \mu ]^2 }
e^{-T E^2} .
\label{eq:h7}
\end{equation}
The $n$ sum, the $\nu$ integral, and the proper time integration can be performed in
closed form. The details of this calculation are presented in
appendix~\ref{app:int}. The final result is:
\begin{eqnarray}
\Gamma_{\beta,\mu} &=& \Gamma_{\infty ,0}
\label{eq:h9} \\
&-& \frac{\Nf d_\gamma}{\beta }  \int d^Dx \int_0^\infty dE \, E \, \rho _0(E) \,
\left[ \ln \left( 1 + e^{-\beta (E-\mu)} \right)
\, + \, \ln \left( 1 + e^{-\beta (E+\mu)} \right)  \right] .
\nonumber
\end{eqnarray}
{which we recognize as the standard expression \cite{Kapusta:1989tk}.}
Although this result was derived under the assumption (\ref{eq:h4}),
the closed form (\ref{eq:h9}) is perfectly valid for $\beta  \mu
>\pi $. Hence, we perform an analytic continuation by taking
(\ref{eq:h9}) as the primary expression and consider the arbitrary values
for $\beta  \mu$ in the following.

Let us generalize the above findings to the more general case of a
spatially dependent background field:
\begin{equation}
	\Gamma_{\beta,\mu}
	= \frac{1}{2}
	\int d^Dx
	\frac{\Nf d_\gamma}{(4\pi)^{D/2}}
	\int_{1/\Lambda^2}^\infty\frac{dT}{T^{1+D/2}}
	\biggl(
	\sum_{n=-\infty}^\infty (-1)^n  e^{-\frac{\beta^2 n^2}{4T}}
	e^{ - \mu \beta n}
	\biggr)
        \Bigl\langle
	e^{-\int_0^Td\tau \sigma^2} \Phi[\sigma] \Bigr\rangle.
	\label{eq:h10}
\end{equation}
The key observation is that we can relate the general case (\ref{eq:h10})
to the free case by defining the density of states $\rho (E) $ by
\begin{eqnarray}
\frac{1}{T^{(D-1)/2} } \, \Bigl\langle
	e^{-\int_0^Td\tau \sigma^2} \Phi[\sigma] \Bigr\rangle
   &=& 2 (4\pi )^{(D-1)/2}
\int _0^\infty dE \, E \, \rho (E) \, e^{-TE^2} \, ,
\label{eq:h11}
\end{eqnarray}
The final answer for the effective action $ \Gamma_{\beta,\mu} $ is given by
(\ref{eq:h9}) where we replace $\rho _0$ by $\rho $ from (\ref{eq:h11}).
Note that the left-hand side of (\ref{eq:h11}) is only known numerically
for real values $T$. The quite challenging task is then to
perform the inverse Laplace transform to obtain $\rho (E)$.
We however point out that, depending on the parameters, the
full range of $\rho (E)$ is not needed. Let us consider the
particle density $n_{\beta, \mu }$ for the moment $(V= \int d^Dx )$,
\begin{equation}
	n_{\beta,\mu} = -\frac{1}{V} \frac{ \partial \Gamma_{\beta,\mu}}{
\partial \mu } =
 \Nf d_\gamma  \int_0^\infty dE \, E \, \rho (E) \,
\left[  \frac{1}{e^{\beta (E-\mu)} +1 } \, - \,
\frac{1}{e^{\beta (E+\mu)} +1 }  \right] .
\label{eq:h15}
\end{equation}
For small temperatures, i.e., for $\beta \mu \gg 1$, we obtain the result ({ valid if $\mu$ lies within a band}):
\begin{equation}
n_{\beta,\mu} =
 \Nf d_\gamma  \left\{ \int _0^\mu  dE \, E \, \rho (E) \, + \,
\frac{\pi^2 }{6 \beta ^2 } \frac{d}{dE} \left[ E \rho (E) \right]
\Bigl\vert _{E=\mu } \, + \, {\cal O}(1/\beta ^4) \right\} .
\label{eq:h16}
\end{equation}
In this case, the density of states need only be known for
$E \stackrel{<}{_\sim } \mu $.

\subsection{Crystalline phase on the worldline}
\label{sec:crystal}

The crystalline phase of the GN model extends up to the tricritical point
plotted in Fig.~(\ref{fig:oldpd}) \cite{Thies:2003br}. This point and
therewith some part of the crystalline phase is above the line defined by
$\beta \mu = \pi$, where $\beta \mu > \pi$. Consequently, already the
straight forward evaluation of Eq.~\eqref{eq:Gthermo} allows for studying
translational-symmetry breaking.

We consider several points on a line slightly above the
$\beta \mu = \pi$ line, given by $1.1 \beta \mu = \pi$, which
intersects the crystalline phase between $\mu = 0.63m$ and $\mu = 0.67m$.
For each point, we determine the potential $\sigma$ of the ground state
analytically as described in \cite{Schnetz:2004vr}. Using this potential, we
evaluate Eq.~\eqref{eq:Gren} and Eq.~\eqref{eq:Gthermo} with
worldline numerics and compare the resulting
value of the grand potential per volume with the corresponding value for
the constant field solution.

\FIGURE[t]{
\centering
\includegraphics{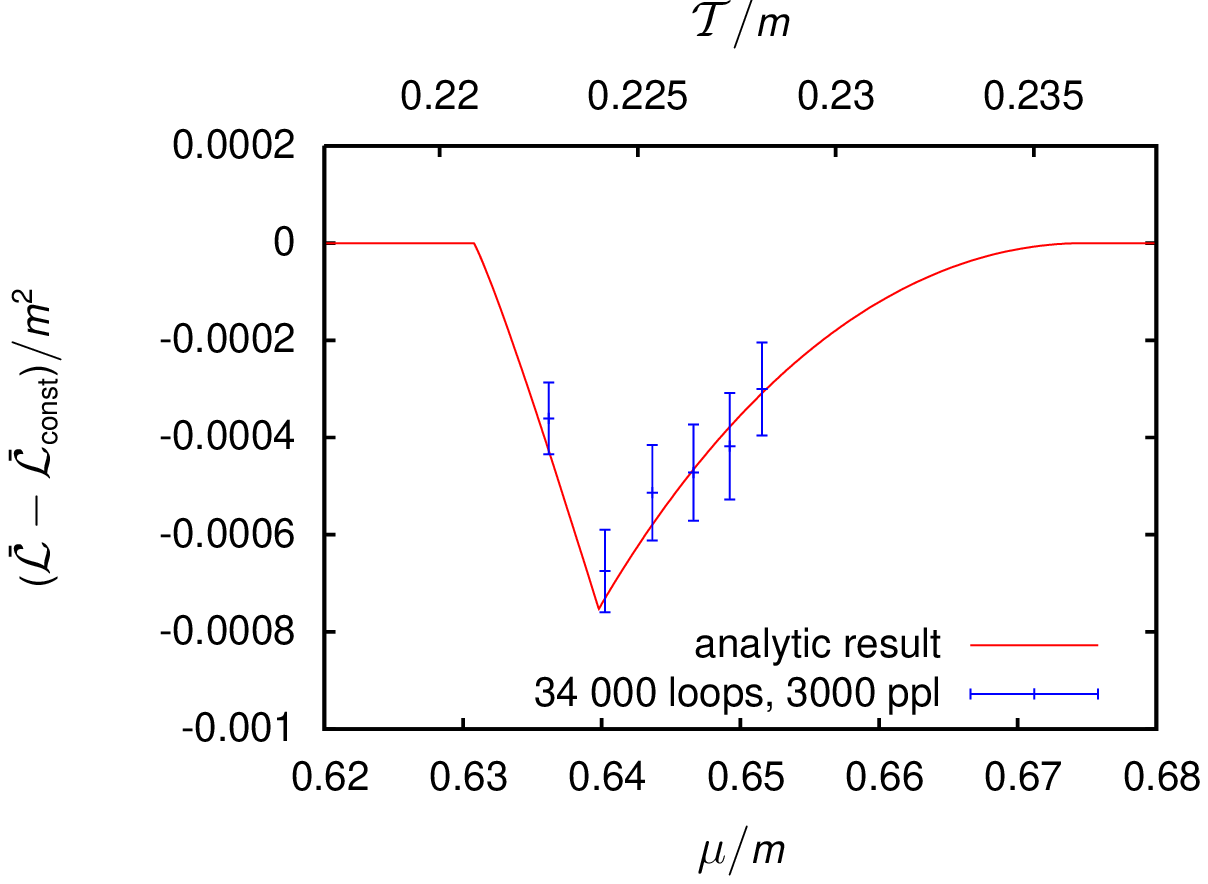}
\caption{
Difference between the grand
potential per volume for the crystal and the homogeneous solution
for $\mathcal T = 1.1 \mu / \pi$. The solid line is the analytic result from
\cite{Schnetz:2004vr}, the
crosses with error bars are worldline numerical values.
The line $\mathcal T = 1.1 \mu / \pi$ crosses the crystal phase between $\mu = 0.63 m$
and $0.67 m$, approximately. Consequently, we obtain negative values in
this range. The numerical values agree with the analytic result and are
significantly below zero.
}
\label{fig:delta_g}
}
Figure~(\ref{fig:delta_g}) shows the difference of these two values
as well as the analytic result according to \cite{Schnetz:2004vr}.
Both results agree nicely. In particular, the worldline numerical result
confirms the thermodynamic preference of the crystal solution over the
homogeneous potential in the range $0.63 m < \mu < 0.67 m$ which is
reflected in negative values of the plotted difference.

\section{Conclusions}

We have demonstrated that the Monte Carlo approach to the worldline
formulation of quantum field theory may be consistently implemented to study
interacting fermionic models such as the Gross-Neveu model.  We have
explained the required formalism and tested numerically the worldline
results in comparison to a number of analytic results for inhomogeneous
condensates, pointing out the advantages and disadvantages of the numerical
approach. In particular, in Section \ref{sec:crystal} we have used this
approach to confirm the existence of the inhomogeneous crystalline phase of
the Gross-Neveu model. While this is a highly non-trivial test, the true
potential of the worldline Monte Carlo approach lies in the possibility of
scaling up to higher dimensions without dramatic numerical problems, and the
possibility of realizing chiral symmetries directly. These issues are under
investigation.

\acknowledgments

KL thanks Simon Hands for helpful discussions on dense fermionic systems.
GD thanks the DOE for support through grant DE-FG02-92ER40716, and the DFG for
support through a Mercator Professor Award held at Heidelberg in 2007, when
this work was begun.  \colH{HG acknowledges support by the DFG under contract Gi
328/1-4 (Emmy-Noether program) and Gi 328/5-1 (Heisenberg-Program).}
The work of KK was supported in part by the National Science Foundation
under Grant No.~PHY05-51164.

\appendix

\section{Fermionic models and continuous chiral symmetry}
\label{app:thirring}

In this work, we have mainly focussed on the the $D=1+1$ dimensional Gross-Neveu
(GN) model defined by the Euclidean fermionic action in (\ref{eq:SF}). The GN model has a discrete chiral
symmetry under $\psi\to \gamma_5 \psi$, rather than a continuous chiral symmetry. In order to
discuss the fate of chiral symmetry in our numerical worldline approach, it is
instructive to study the Thirring model, defined by the action
\begin{equation}
S_{\text{F}}= \int d^Dx
	\biggl(
		-\bar\psi
		\fss\partial \psi
		+ \frac{g^2}{2\Nf}  ( \yb \gamma_\mu \psi )^2
	\biggr), \label{eq:SFT}
\end{equation}
\colH{In addition to global phase and axial rotations,} the Thirring model is
invariant under SU($\Nf$)$_{\text{R}} \times$ SU($\Nf$)$_{\text{L}}$ chiral
transformations. By means of a Hubbard-Stratonovich transformation, the model
can be rewritten with the aid of a vector boson $V_\mu$
\begin{equation}
S_{\text{FB}}= \int d^D x \bigg[ \frac{\Nf}{2 g^2}V_\mu^2
    -\yb \fss{\partial} \psi - i V_\mu
  \yb \gamma_\mu \psi  \bigg], \label{eq:SFBT}
\end{equation}
where $V_\mu$ is invariant under chiral transformations. In the large-$\Nf$
limit, the effective action upon integrating out the fermion fluctuations
yields
\begin{equation}
  \Gamma[V_\mu]= \int d^D x  \frac{\Nf}{2 g^2}  V_\mu^2
  - \Nf\, \Tr \ln (-\fss{\partial} - i  \fsV{V} ), \quad
  \Nf\to\infty. \label{eq:Gam0T}
\end{equation}
As in the Gross-Neveu case, the resulting Dirac operator $\mathcal D = i
\fss{\partial} -\fsV{V}$ has a $\gamma_5$ hermiticity,
$\mathcal{D}^\dagger=\gamma_5 \mathcal{D} \gamma_5$, implying that $\Gamma$
can be written as
\begin{equation}
\Gamma[V_\mu]= \int d^D x  \frac{\Nf}{2 g^2} V_\mu^2
 -\frac{\Nf}{2} \Tr \ln \left(-D[V]^2 +
  \frac{1}{2} \sigma_{\mu\nu} V_{\mu\nu} \right), \quad \Nf\to\infty, \label{eq:Gam1T}
\end{equation}
where $D[V]=\partial_\mu + i V_\mu$, $\sigma_{\mu\nu}=\frac{i}{2}
[\gamma_\mu,\gamma_\nu]$, and the vector field strength
$V_{\mu\nu}=\partial_\mu V_\nu -\partial_\nu V_\mu$. By virtue of the $D=1+1$
dimensional Dirac algebra, the spin-field coupling (Pauli term) can also be
expressed as $\sigma_{\mu\nu} V_{\mu\nu}=\gamma_5 \epsilon_{\mu\nu}
V_{\mu\nu}\equiv \gamma_5 \widetilde{V}$. Here, we prefer to work with the
representation given in \Eqref{eq:Gam1T} which generalizes straightforwardly to
higher dimensions\footnote{Odd dimensions require reducible representations of
  the Dirac algebra in order to maintain $\gamma_5$ hermiticity of the Dirac
  operator.}. Note that the fermionic contribution exhibits an accidental
local gauge symmetry, which is, however, not respected by the classical action
$\sim V_\mu^2$.

Incidentally, the 4-fermion vector interaction of the Thirring model is the
only SU($\Nf$)$_{\text{R}}$ $\times$ SU($\Nf$)$_{\text{L}}$ chirally invariant
and parity-even interaction which contributes to leading-order at
large-$\Nf$ in $D=2$. In higher dimensions, also independent axial-vector
interactions exist which, in $D=2$, are isomorphic to the vector interaction.

The worldline representation for the Thirring model is well known from the
analogous  QED case. Concentrating on the contributions from the fermion
fluctuations, we obtain
\begin{eqnarray}
\Delta\Gamma[V]&:=&-\frac{\Nf}{2} \Tr \ln \left(-D[V]^2 +
  \frac{1}{2}\sigma_{\mu\nu} F_{\mu\nu}\right) \nonumber\\
&=&\frac{\Nf}{2} \frac{1}{(4\pi)^{D/2}}\, d_\gamma
\int\limits_{1/\Lambda^2}^\infty \frac{dT}{T^{D/2+1}}  \!\!
\int\limits_{x(0)=x(T)} \!\!\mathcal{D} x \, e^{-\frac{1}{4} \int_0^T
  d\tau \, \dot{x}^2(\tau)} \, e^{-i \oint dx_\mu V_\mu } \Phi[V],
\label{eq:DelG2T}
\end{eqnarray}
where the spin factor this time reads
\begin{equation}
\Phi[V]=\frac{1}{d_\gamma}\, \tr_\gamma \; {\cal P} \, \hbox{exp} \left(
  -\frac{1}{2} \, \int
 _0^T d\tau \, \sigma_{\mu\nu} V_{\mu\nu}  \right) .
\end{equation}

\section{Exact resolvents and propagators for $(1+1)$-dim.
Gross-Neveu  model
\label{app:oldA} }

In this Appendix, we present the derivation of the exact expressions for
$\langle e^{-\int_0^T d\tau\, V_\pm}\rangle$ discussed in Section
\ref{inhomog}, where they were also compared to our numerical worldline
results. We compute, for various forms of the static condensate function
$\sigma(x)$, the worldline propagators
\bea
K_\pm(x; T) \equiv \langle x\, | e^{-H_\pm \, T}\, |\, x\rangle \equiv
\frac{1}{\sqrt{4\pi\, T}} \, \langle e^{-\int_0^T d\tau\, V_\pm}\rangle \quad ,
\label{propagator}
\eea
where the Schr\"odinger-like Hamiltonian operators $H_\pm$ are
\bea
H_\pm\equiv -\partial_x^2+V_\pm(x)=-\partial_x^2+\sigma^2(x)\pm \sigma^\prime(x)\quad .
\label{hpm}
\eea
Our strategy is to consider the associated "diagonal resolvents" [i.e.,
coincident-point Green's functions], which are Laplace transforms of the
propagators:
\bea
R_\pm(x; -\lambda)\equiv \langle x\, | \frac{1}{H_\pm +\lambda}\,| x\rangle
=\int_0^\infty dT\, e^{-\lambda T}  \langle x\, | e^{-H_\pm \, T}\, |\,
x\rangle\ \quad .
\label{resolvent2}
\eea
For {\it any} static condensate function $\sigma(x)$, the resolvent must
satisfy the Gel'fand-Dik'ii equation \cite{joshua}
\bea
-2R\, R^{\prime\prime}+\left(R^\prime\right)^2+4R^2\left(V+\lambda\right)=1\quad ,
\label{gd1}
\eea
or its third-order linear form [obtained by differentiating (\ref{gd1})]
\bea
R^{\prime\prime\prime}-4 R^\prime\left(V+\lambda\right)-2 R V^\prime=0
\label{gd2}
\eea
This identity follows directly from the fact that the Green's function in 1
dimension can be written in terms of the product of two independent solutions,
and the product of two solutions necessarily satisfies such an equation. For
details, see \cite{joshua}.

For certain special forms of static condensate $\sigma(x)$, these
Gel'fand-Dik'ii equations are analytically soluble. Then $K_\pm(x; T)$ is
obtained as the inverse Laplace transform of $R_\pm(T; -\lambda)$. Let us
consider several examples:

\subsection{Homogeneous condensate}
With a homogeneous condensate $\sigma(x)=m$, we have $V_\pm=m^2$, so that
solving (\ref{gd1})--(\ref{gd2}) leads to
\bea
R_\pm(x; -\lambda)=\frac{1}{2\sqrt{m^2+\lambda}} \quad .
\label{free-resolvent}
\eea
It is straightforward to take the inverse Laplace transform to arrive at the
free propagator in (\ref{eq:free-propagator}).

\subsection{Single kink condensate}

Consider the inhomogeneous
kink-like condensate $ \sigma(x)=A\,\tanh(A\, x) $, for which
\bea
V_\pm =\begin{cases}
A^2\cr
A^2(1-2\,{\rm sech}^2(A\, x)).
\end{cases}
\label{potentials-kink2}
\eea
Then $R_\pm(x;-\lambda)$ can be obtained
from (\ref{gd1})--(\ref{gd2})  by substituting the {\it ansatz} form
\bea
R_\pm(x;-\lambda)=\alpha_\pm(\lambda)+\beta_\pm(\lambda) \,{\rm sech}^2(A\, x)\quad ,
\label{gd-ansatz}
\eea
and solving algebraically for the [$\lambda$-dependent] coefficients
$\alpha_\pm$ and $\beta_\pm$. A simple calculation leads to
\bea
R_\pm(x;-\lambda)=
\begin{cases}
\frac{1}{2\sqrt{A^2+\lambda}} \cr
\frac{1}{2\sqrt{A^2+\lambda}}\left(1+\frac{A^2}{\lambda}
{\rm sech}^2(A\, x)
\right)
\end{cases}.
\label{kink-resolvent}
\eea
The corresponding propagators follow from the inverse Laplace transform, and
are presented in (\ref{kink-propagator}).

\subsection{Kink-antikink pair condensate}

Consider the inhomogeneous condensate corresponding to a kink-antikink pair
\bea
\sigma(x)&=&A\left(\coth(b)+\left[\tanh(A\, x)-\tanh(A\, x+b)\right]\right) \nn
&=& A\left(\left[\coth(b)-\tanh(b)\right]+\tanh(b)\tanh(A\, x)\tanh(A\, x+b)\right)\quad .
\label{kkbar-sigma}
\eea
The corresponding potentials are
\bea
V_\pm &=&A^2
\left( \coth^2(b)
-2 \left\{
\begin{matrix}
{\rm sech}^2(A\, x+b) \\
{\rm sech}^2(A\, x)
\end{matrix}
\right\}
\right) \quad .
\label{kkbar-potential}
\eea
The resolvents can be derived by similar ansatze to (\ref{gd-ansatz}), using
${\rm sech}^2(A\, x+b)$ for $V_+$, and ${\rm sech}^2(A\, x)$ for $V_-$. This
leads straightforwardly to
\bea
R_\pm(x; -\lambda)=\frac{1}{2\sqrt{A^2
    \coth^2(b)+\lambda}}\left(1+\frac{A^2}{\frac{A^2}{\sinh^2(b)}+\lambda}
 \left\{
\begin{matrix}
{\rm sech}^2(A\, x+b) \\
{\rm sech}^2(A\, x)
\end{matrix}
\right\}
\right)
\label{kkbar-resolvent}
\eea
The propagators again follow from the inverse Laplace transform, and are
presented in (\ref{kkbar-propagator}).

\subsection{Periodic array of kink-antikink pairs}

Consider the condensate formed by a periodic array of kink-antikink pairs
\bea
\sigma(x)&=&A\left(\left[ Z(b; \nu)+\frac{{\rm cn}(b; \nu){\rm dn}(b;
      \nu)}{{\rm sn}(b; \nu)}\right]+\left[Z(A\, x; \nu)-Z(A\, x+b;
    \nu)\right]\right)\nn
&=&A\left(\frac{{\rm cn}(b; \nu){\rm dn}(b; \nu)}{{\rm sn}(b; \nu)}+\nu\, {\rm
    sn}(b; \nu)\, {\rm sn}(A\, x; \nu)\, {\rm sn}(A\, x+b; \nu)\right) \quad
.
\label{periodic-sigma}
\eea
The potentials corresponding to the periodic condensate (\ref{periodic-sigma}) are
\bea
V_\pm=A^2\left( \frac{1}{{\rm sn}^2(b; \nu)}-1+\nu -2\nu  \left\{
\begin{matrix}
{\rm cn}^2(A\, x+b; \nu) \\
{\rm cn}^2(A\, x; \nu)
\end{matrix}
\right\}
\right) \quad .
\label{periodic-potential}
\eea
The resolvents for $V_\pm$ can be obtained by substituting ansatze as in
(\ref{gd-ansatz}), but with ${\rm sech}^2$ replaced by ${\rm cn}^2$, and with
the argument shifted by $b$ in the case of $V_+$. Straightforward algebra
yields
\bea
R_\pm(x; -\lambda)=&&\nonumber\\
&&\hskip -30pt \frac{\sqrt{A^2 {\rm ds}^2(b; \nu)+\lambda}}{2\sqrt{A^2 {\rm
      cs}^2(b; \nu)+\lambda}\sqrt{A^2 {\rm ns}^2(b; \nu)+\lambda}}
\left(1+\frac{A^2\, \nu}{A^2 {\rm ds}^2(b; \nu)+\lambda}
 \left\{
\begin{matrix}
{\rm cn}^2(A\, x+b; \nu) \\
{\rm cn}^2(A\, x; \nu)
\end{matrix}
\right\}
\right)\,\, .\nn
\label{periodic-resolvent}
\eea
The propagators again follow from the inverse Laplace transform, and are
presented in (\ref{periodic-propagator}). The kink crystal case, where the
antikinks lie at the edges of the period, is obtained by the simple
substitution $b\to{\rm K}(\nu)$.

\subsection{Comparison with heat kernel expansion}
\label{app:hke}

It is instructive to compare the exact transition amplitudes, or heat kernels,
not only with our numerical worldline loop results, but also with the
approximate small $T$ heat kernel expansions
\cite{Gilkey:1995mj,Kirsten:2001wz,Vassilevich:2003xt}. In general, the
transition amplitude/heat kernel has an asymptotic small $T$ expansion
\bea
K(x; T)\sim\frac{1}{\sqrt{4\pi\, T}}\sum_{n=0}^\infty b_n(x)\, T^n\quad ,
\label{heat-kernel}
\eea
where the functionals $b_n(x)$ are expressed in terms of the potential and its
derivatives, and are given by:
\bea
b_0(x)&=&1\nonumber\\
b_1(x)&=&-V \nonumber\\
b_2(x)&=&\frac{1}{2}\left(V^2-\frac{1}{3}V^{\prime\prime}\right) \nonumber\\
b_3(x)&=&-\frac{1}{6}\left(V^3-\frac{1}{2}(V^\prime)^2-V\, V^{\prime\prime}+\frac{1}{10} V^{\prime\prime\prime\prime}\right)\nonumber\\
&\vdots& \eea Defining $b_n(x)=\frac{2^{n+1}}{(2n-1)!!}\,r_n(x)$, the $r_n$
are given recursively by : \bea r_n=\frac{1}{2}\sum_{k=0}^{n-2}r_k\,
r_{n-k-1}^{\prime\prime}-\frac{1}{4}\sum_{k=1}^{n-2}r_k^\prime\,
r_{n-k-1}^\prime-V\sum_{k=0}^{n-1}r_k\,r_{n-k-1}-\sum_{k=1}^{n-1}r_k\,
r_{n-k}\quad ,
\label{heat}
\eea
with $r_0=\frac{1}{2}$ and $r_1=-\frac{1}{4}V$.

For example, for the single-kink condensate, with $V_-=A^2\left(1-2\, {\rm
    sech}^2(A\, x)\right)$, the heat kernel expansion (\ref{heat-kernel}) is
\bea
K_-(x; T)&=&\frac{1}{\sqrt{4\pi\, T}} \left\{ 1-A^2 T\left[1-2\, {\rm sech}^2(A\, x)\right]
+\frac{1}{2}A^4\, T^2 \left[ 1-\frac{4}{3}\, {\rm sech}^2(A\, x)\right] \right .\nonumber\\
&&\left. \hskip 2cm -\frac{1}{6}A^6\, T^3 \left[1-\frac{6}{5}\, {\rm sech}^2(A\, x)\right] +\dots
\right\}\quad ,
\eea
in agreement with the small $T$ expansion of the exact result in
(\ref{kink-propagator}). This comparison is made in the first plot of Figure \ref{fig:overlap}.
Similarly, for the periodic kink-antikink condensate, with $V_-(x)$ as in
(\ref{periodic-potential}), the heat kernel expansion (\ref{heat-kernel}) is
\bea K_-(x; T)&=&\frac{1}{\sqrt{4\pi\, T}} \left\{ 1-A^2\,T
  \left[\nu-1+\frac{1}{{\rm sn}^2(b; \nu)}-2\,\nu\,
    \text{cn}^2\left(A\, x;\nu\right) \right]\right.\nonumber\\
&&  \quad+\frac{1}{2} A^4\,
  T^2\left[\frac{(1-\nu)(\nu+3)}{3}+\frac{2(\nu-1)}{{\rm sn}^2(b; \nu)}+
    \frac{1}{{\rm sn}^4(b; \nu)} \right.\nonumber \\
&&\hspace{2cm}  \left.\left.-\left(\frac{4\nu}{{\rm sn}^2(b;
        \nu)}-\frac{4}{3}\nu(\nu+1)\right) \text{cn}^2\left(A\, x;\nu\right)
  \right] +\dots \right\},
\eea
in agreement with the small
$T$ expansion of the exact result in (\ref{periodic-propagator}). This
comparison is made in Figure \ref{fig:pathint}.

\section{Effective action for the homogeneous case \label{app:int}}

This section provides detailed information on the evaluation of the integrals
and sums which appear in the effective action (\ref{eq:h7}) for a
homogeneous background field. We firstly focus on the contributions
from the terms with $n \not=0$:
\begin{equation}
I(E) =
\frac{1}{\beta } \sum_{n\not=0}
\int_{0}^\infty\frac{dT}{T}
\int d\nu \;
e^{i 2 \pi n \, \nu } \, e^{- T [\frac{2\pi}{\beta } (\nu +\frac{1}{2}) - i
  \mu ]^2 }
e^{-T E^2} .
\label{eq:hh1}
\end{equation}
Let us first perform the propertime integration. Using the identity
\begin{equation}
2 \int _E^\infty dx \; x \; e^{-Tx^2} = \frac{1}{T} e^{-T E^2} \; ,
\label{eq:hh2}
\end{equation}
the $T$ integration can be be performed:
\begin{equation}
I(E) = \frac{2}{\beta } \int _E^\infty dx \, x \, \sum_{n\not=0} \int d\nu \;
e^{i 2\pi n \nu } \; \frac{1}{ [ \frac{2\pi}{\beta } (\nu +\frac{1}{2})
-i \mu ]^2 + x^2 } .
\label{eq:hh3}
\end{equation}
The $\nu $ integration requires to distinguish the cases
$x>\mu $ and $x<\mu$, but can otherwise be straightforwardly done
using the Cauchy integral theorem:
\begin{equation}
I(E) =
\int _E^\infty dx \, \sum_{n=1}^\infty (-1)^n
\Bigl[ e^{-\beta n (x+\mu)} \, + \, \hbox{sign} (x-\mu) \,
e^{-\beta n \vert x-\mu \vert } \Bigr] .
\label{eq:hh4}
\end{equation}
The geometric series in $n$ can easily be summed:
\begin{equation}
I(E) =
 -  \int _E^\infty dx \,
\left[ \frac{e^{-\beta (x+\mu)} }{1 + e^{-\beta (x+\mu)} }
\, + \, \hbox{sign} (x-\mu) \,
\frac{e^{-\beta \vert x-\mu\vert} }{1 + e^{-\beta \vert x-\mu \vert } } \right] .
\label{eq:hh5}
\end{equation}
Using the identity
$$
\frac{e^{-\beta (\mu-x) } }{1 + e^{-\beta  (\mu-x) } }
= 1 - \frac{e^{ - \beta (x-\mu) } }{1 + e^{- \beta  (x-\mu) } } ,
$$
the expression (\ref{eq:hh5}) can be written as
\begin{equation}
I(E) =
 -  \int _E^\infty dx \,
\left[ \frac{e^{-\beta (x+\mu)} }{1 + e^{-\beta (x+\mu)} }
\, - \, \theta (\mu-x) \, + \,
\frac{e^{-\beta (x-\mu) } }{1 + e^{-\beta (x-\mu) } } \right] .
\label{eq:hh6}
\end{equation}
Let us now consider the contribution with zero winding $n=0$ in \Eqref{eq:h7}:
\begin{equation}
I_0(E,\mu) =
\frac{1}{\beta }
\int_{1/\Lambda ^2}^\infty\frac{dT}{T}
\int d\nu \; e^{- T [\frac{2\pi}{\beta } (\nu +\frac{1}{2}) - i
  \mu ]^2 }
e^{-T E^2} .
\label{eq:hh7}
\end{equation}
It is convenient to decompose the latter expression as
\begin{eqnarray}
I_0(E,\mu) &=& \Delta I_0(E,\mu) + I_0(E,0) ,
\label{eq:hh8} \\
\Delta I_0(E,\mu) &=& \frac{1}{\beta }
\int_{0}^\infty\frac{dT}{T} \; e^{-T E^2}
\int d\nu \;
\left[ e^{- T [\frac{2\pi}{\beta } (\nu +\frac{1}{2}) - i  \mu ]^2 } -
e^{- T [\frac{2\pi}{\beta } (\nu +\frac{1}{2}) ]^2 } \right] .
\label{eq:hh9}
\end{eqnarray}
Note that the latter expression is UV finite. We have therefore removed
the regulator by taking the limit $\Lambda \to \infty $.
A direct calculation of $\Delta I_0(E,\mu) $ along the same lines outlined
above yields
\begin{equation}
\Delta I_0(E,\mu)  = -  \int _E^\infty dx \, \theta(\mu - x) .
\label{eq:hh10}
\end{equation}
Thus, combining (\ref{eq:hh6}) and (\ref{eq:hh7}) gives
\begin{equation}
I(E) + I_0(E,\mu)  = I_0(E,0) -  \int _E^\infty dx \,
\left[ \frac{e^{-\beta (x+\mu)} }{1 + e^{-\beta (x+\mu)} }
\, + \, \frac{e^{-\beta (x-\mu) } }{1 + e^{-\beta (x-\mu) } } \right] .
\label{eq:hh11}
\end{equation}
The final integration can easily be done leaving us with
\begin{equation}
I(E) + I_0(E,\mu)  = I_0(E,0) - \frac{1}{\beta }
 \left[ \ln \left( 1 + e^{-\beta (E-\mu)} \right)
\, + \, \ln \left( 1 + e^{-\beta (E+\mu)} \right)  \right]  .
\label{eq:hh12}
\end{equation}


\begin{thebibliography}{999}


\bibitem{Stephanov:2007fk}
  M.~A.~Stephanov,
  ``QCD phase diagram: An overview,''
  PoS {\bf LAT2006}, 024 (2006)
  [arXiv:hep-lat/0701002].
  %%CITATION = POSCI,LAT2006,024;%%

  %\cite{Choe:2002mt}
\bibitem{Choe:2002mt}
  S.~Choe {\it et al.},
  ``Responses of hadrons to the chemical potential at finite temperature,''
  Phys.\ Rev.\  D {\bf 65}, 054501 (2002).
  %%CITATION = PHRVA,D65,054501;%%

%\cite{Allton:2002zi}
\bibitem{Allton:2002zi}
  C.~R.~Allton {\it et al.},
  ``The QCD thermal phase transition in the presence of a small chemical
  potential,''
  Phys.\ Rev.\  D {\bf 66}, 074507 (2002)
  [hep-lat/0204010].
  %%CITATION = PHRVA,D66,074507;%%

%\cite{Ejiri:2003dc}
\bibitem{Ejiri:2003dc}
  S.~Ejiri, C.~R.~Allton, S.~J.~Hands, O.~Kaczmarek, F.~Karsch,
  E.~Laermann and C.~Schmidt,
  ``Study of QCD thermodynamics at finite density by Taylor expansion,''
  Prog.\ Theor.\ Phys.\ Suppl.\  {\bf 153}, 118 (2004)
  [hep-lat/0312006].
  %%CITATION = PTPSA,153,118;%%


%\cite{Alford:1998sd}
\bibitem{Alford:1998sd}
  M.~G.~Alford, A.~Kapustin and F.~Wilczek,
  ``Imaginary chemical potential and finite fermion density on the lattice,''
  Phys.\ Rev.\  D {\bf 59}, 054502 (1999)
  [hep-lat/9807039].
  %%CITATION = PHRVA,D59,054502;%%

%\cite{de Forcrand:2002ci}
\bibitem{de Forcrand:2002ci}
  P.~de Forcrand and O.~Philipsen,
  ``The QCD phase diagram for small densities from imaginary chemical
  potential,''
  Nucl.\ Phys.\  B {\bf 642}, 290 (2002)
  [hep-lat/0205016].
  %%CITATION = NUPHA,B642,290;%%

%\cite{D'Elia:2002gd}
\bibitem{D'Elia:2002gd}
  M.~D'Elia and M.~P.~Lombardo,
  ``Finite density QCD via imaginary chemical potential,''
  Phys.\ Rev.\  D {\bf 67}, 014505 (2003)
  [hep-lat/0209146].
  %%CITATION = PHRVA,D67,014505;%%

%\cite{Fodor:2001au}
\bibitem{Fodor:2001au}
  Z.~Fodor and S.~D.~Katz,
  ``A new method to study lattice QCD at finite temperature and chemical
  potential,''
  Phys.\ Lett.\  B {\bf 534}, 87 (2002)
  [hep-lat/0104001]; %.
  %%CITATION = PHLTA,B534,87;%%
%
%\cite{Fodor:2001pe}
%\bibitem{Fodor:2001pe}
%  Z.~Fodor and S.~D.~Katz,
  ``Lattice determination of the critical point of QCD at finite T and mu,''
  JHEP {\bf 0203}, 014 (2002)
  [hep-lat/0106002]; %.
  %%CITATION = JHEPA,0203,014;%%
%
%\cite{Fodor:2004nz}
%\bibitem{Fodor:2004nz}
%  Z.~Fodor and S.~D.~Katz,
  ``Critical point of QCD at finite T and mu, lattice results for physical
  quark masses,''
  JHEP {\bf 0404}, 050 (2004)
  [hep-lat/0402006].
  %%CITATION = JHEPA,0404,050;%%

%\cite{Hands:1999md}
\bibitem{Hands:1999md}
  S.~Hands, J.~B.~Kogut, M.~P.~Lombardo and S.~E.~Morrison,
  ``Symmetries and spectrum of SU(2) lattice gauge theory at finite  chemical
  potential,''
  Nucl.\ Phys.\  B {\bf 558}, 327 (1999)
  [hep-lat/9902034].
  %%CITATION = NUPHA,B558,327;%%

%\cite{Kogut:2002cm}
\bibitem{Kogut:2002cm}
  J.~B.~Kogut, D.~Toublan and D.~K.~Sinclair,
  ``The phase diagram of four flavor SU(2) lattice gauge theory at nonzero
  chemical potential and temperature,''
  Nucl.\ Phys.\  B {\bf 642}, 181 (2002)
  [hep-lat/0205019].
  %%CITATION = NUPHA,B642,181;%%

%\cite{Hands:2006ve}
\bibitem{Hands:2006ve}
  S.~Hands, S.~Kim and J.~I.~Skullerud,
  ``Deconfinement in dense 2-color QCD,''
  Eur.\ Phys.\ J.\  C {\bf 48}, 193 (2006)
  [hep-lat/0604004].
  %%CITATION = EPHJA,C48,193;%%

%\cite{Bailin:1983bm}
\bibitem{Bailin:1983bm}
  D.~Bailin and A.~Love,
  ``Superfluidity And Superconductivity In Relativistic Fermion Systems,''
  Phys.\ Rept.\  {\bf 107}, 325 (1984).
  %%CITATION = PRPLC,107,325;%%

%\cite{Alford:1997zt}
\bibitem{Alford:1997zt}
  M.~G.~Alford, K.~Rajagopal and F.~Wilczek,
  ``QCD at finite baryon density: Nucleon droplets and color
  superconductivity,''
  Phys.\ Lett.\  B {\bf 422}, 247 (1998)
  [hep-ph/9711395].
  %%CITATION = PHLTA,B422,247;%%

%\cite{Schafer:1999jg}
\bibitem{Schafer:1999jg}
  T.~Schafer and F.~Wilczek,
  ``Superconductivity from perturbative one-gluon exchange in high density
  quark matter,''
  Phys.\ Rev.\  D {\bf 60}, 114033 (1999)
  [hep-ph/9906512].
  %%CITATION = PHRVA,D60,114033;%%

%\cite{Pisarski:1999tv}
\bibitem{Pisarski:1999tv}
  R.~D.~Pisarski and D.~H.~Rischke,
  ``Color superconductivity in weak coupling,''
  Phys.\ Rev.\  D {\bf 61}, 074017 (2000)
  [nucl-th/9910056].
  %%CITATION = PHRVA,D61,074017;%%


  \bibitem{Alford:2007xm}
  M.~G.~Alford, A.~Schmitt, K.~Rajagopal and T.~Schafer,
  ``Color superconductivity in dense quark matter,''
  Rev.\ Mod.\ Phys.\  {\bf 80}, 1455 (2008)
  [arXiv:0709.4635 [hep-ph]].
  %%CITATION = RMPHA,80,1455;%%

  \bibitem{Alford:2000ze}
  M.~G.~Alford, J.~A.~Bowers and K.~Rajagopal,
  ``Crystalline color superconductivity,''
  Phys.\ Rev.\  D {\bf 63}, 074016 (2001)
  [arXiv:hep-ph/0008208].
  %%CITATION = PHRVA,D63,074016;%%

  \bibitem{Rajagopal:2000wf}
  K.~Rajagopal and F.~Wilczek,
  ``The condensed matter physics of QCD,''
  arXiv:hep-ph/0011333.
  %%CITATION = HEP-PH/0011333;%%

  \bibitem{Bowers:2002xr}
  J.~A.~Bowers and K.~Rajagopal,
  ``The crystallography of color superconductivity,''
  Phys.\ Rev.\  D {\bf 66}, 065002 (2002)
  [arXiv:hep-ph/0204079].
  %%CITATION = PHRVA,D66,065002;%%

  \bibitem{casalbuoni}
  R.~Casalbuoni and G.~Nardulli,
  ``Inhomogeneous superconductivity in condensed matter and QCD,''
  Rev.\ Mod.\ Phys.\  {\bf 76}, 263 (2004)
  [arXiv:hep-ph/0305069].
  %%CITATION = RMPHA,76,263;%%


   \bibitem{klebanov}
  I.~R.~Klebanov,
  ``Nuclear Matter In The Skyrme Model,''
  Nucl.\ Phys.\  B {\bf 262}, 133 (1985).
  %%CITATION = NUPHA,B262,133;%%

   \bibitem{goldhaber}
  A.~S.~Goldhaber and N.~S.~Manton,
  ``Maximal Symmetry Of The Skyrme Crystal,''
  Phys.\ Lett.\  B {\bf 198}, 231 (1987).
  %%CITATION = PHLTA,B198,231;%%

  \bibitem{jackson}
  A.~D.~Jackson and J.~J.~M.~Verbaarschot,
  ``Phase structure of the Skyrme model,''
  Nucl.\ Phys.\  A {\bf 484}, 419 (1988);
  %%CITATION = NUPHA,A484,419;%%
  L.~Castillejo, P.~S.~J.~Jones, A.~D.~Jackson, J.~J.~M.~Verbaarschot and A.~Jackson,
  ``Dense Skyrmion Systems,''
  Nucl.\ Phys.\  A {\bf 501}, 801 (1989).
  %%CITATION = NUPHA,A501,801;%%

  \bibitem{manton}
  N.~S.~Manton and P.~M.~Sutcliffe,
  ``Skyrme crystal from a twisted instanton on a four torus,''
  Phys.\ Lett.\  B {\bf 342}, 196 (1995)
  [arXiv:hep-th/9409182].
  %%CITATION = PHLTA,B342,196;%%

    %\cite{Rosenstein:1990nm}
  \bibitem{Rosenstein:1990nm}
  B.~Rosenstein, B.~Warr and S.~H.~Park,
  ``Dynamical symmetry breaking in four Fermi interaction models,''
  Phys.\ Rept.\  {\bf 205}, 59 (1991).
  %%CITATION = PRPLC,205,59;%%



  \bibitem{hands}
  S.~Hands, A.~Kocic and J.~B.~Kogut,
  ``The Four Fermi Model In Three-Dimensions At Nonzero Density And Temperature,''
  Nucl.\ Phys.\  B {\bf 390}, 355 (1993)
  [arXiv:hep-lat/9206024].
  %%CITATION = NUPHA,B390,355;%%





%\cite{Karsch:1986hm}
\bibitem{Karsch:1986hm}
  F.~Karsch, J.~B.~Kogut and H.~W.~Wyld,
  ``The Gross-Neveu Model At Finite Temperature And Density,''
  Nucl.\ Phys.\  B {\bf 280}, 289 (1987).
  %%CITATION = NUPHA,B280,289;%%


\bibitem{Thies:2003br}
M.~Thies,
  ``Analytical solution of the Gross-Neveu model at finite density,''
  Phys.\ Rev.\  D {\bf 69}, 067703 (2004)
  [arXiv:hep-th/0308164];
  %%CITATION = PHRVA,D69,067703;%%
  M.~Thies and K.~Urlichs,
  ``Revised phase diagram of the Gross-Neveu model,''
  Phys.\ Rev.\  D {\bf 67}, 125015 (2003)
  [hep-th/0302092].
  %%CITATION = PHRVA,D67,125015;%%

  %\cite{Thies:2006ti}
  \bibitem{Thies:2006ti}
  M.~Thies,
  ``From relativistic quantum fields to condensed matter and back again:
  Updating the Gross-Neveu phase diagram,''
  J.\ Phys.\ A  {\bf 39}, 12707 (2006)
  [hep-th/0601049]; %.
  %%CITATION = JPAGB,A39,12707;%%

  %\cite{Schnetz:2004vr}
\bibitem{Schnetz:2004vr}
  O.~Schnetz, M.~Thies and K.~Urlichs,
  ``Phase diagram of the Gross-Neveu model: Exact results and condensed  matter
  precursors,''
  Annals Phys.\  {\bf 314} (2004) 425
  [arXiv:hep-th/0402014].
  %%CITATION = APNYA,314,425;%%


  \bibitem{Basar:2008im}
  G.~Basar and G.~V.~Dunne,
  ``Self-consistent crystalline condensate in chiral Gross-Neveu and
  Bogoliubov-de Gennes systems,''
  Phys.\ Rev.\ Lett.\  {\bf 100}, 200404 (2008)
  [arXiv:0803.1501 [hep-th]];
  %%CITATION = PRLTA,100,200404;%%
%G.~Basar and G.~V.~Dunne,
  ``A Twisted Kink Crystal in the Chiral Gross-Neveu model,''
  Phys.\ Rev.\  D {\bf 78}, 065022 (2008)
  [arXiv:0806.2659 [hep-th]];
  %%CITATION = PHRVA,D78,065022;%%
G.~Basar, G.~V.~Dunne and M.~Thies,
  ``Inhomogeneous Condensates in the Thermodynamics of the Chiral $NJL_2$
  model,''
  arXiv:0903.1868 [hep-th].
  %%CITATION = ARXIV:0903.1868;%%

  \bibitem{deForcrand:2006zz}
  P.~de Forcrand and U.~Wenger,
  ``New baryon matter in the lattice Gross-Neveu model,''
  PoS {\bf LAT2006}, 152 (2006)
  [arXiv:hep-lat/0610117].
  %%CITATION = POSCI,LAT2006,152;%%

    \bibitem{bringoltz}
  B.~Bringoltz,
  ``Chiral crystals in strong-coupling lattice QCD at nonzero chemical potential,''
  JHEP {\bf 0703}, 016 (2007)
  [arXiv:hep-lat/0612010];
  %%CITATION = JHEPA,0703,016;%%
``Solving two-dimensional large-N QCD with a nonzero density of baryons and
  arbitrary quark mass,''
  arXiv:0901.4035 [hep-lat].
  %%CITATION = ARXIV:0901.4035;%%

  \bibitem{Nickel:2008ng}
  D.~Nickel and M.~Buballa,
  ``Solitonic ground states in (color-) superconductivity,''
  arXiv:0811.2400 [hep-ph];
  %%CITATION = ARXIV:0811.2400;%%
  D.~Nickel,
  ``How many phases meet at the chiral critical point?,''
  arXiv:0902.1778 [hep-ph].
  %%CITATION = ARXIV:0902.1778;%%




%\cite{Nielsen:1980rz}
\bibitem{Nielsen:1980rz}
  H.~B.~Nielsen and M.~Ninomiya,
  ``Absence Of Neutrinos On A Lattice. 1. Proof By Homotopy Theory,''
  Nucl.\ Phys.\  B {\bf 185}, 20 (1981)
  [Erratum-ibid.\  B {\bf 195}, 541 (1982)].
  %%CITATION = NUPHA,B185,20;%%

%\cite{Kogut:1974ag}
\bibitem{Kogut:1974ag}
  J.~B.~Kogut and L.~Susskind,
  ``Hamiltonian Formulation Of Wilson's Lattice Gauge Theories,''
  Phys.\ Rev.\  D {\bf 11}, 395 (1975).
  %%CITATION = PHRVA,D11,395;%%

%\cite{Kaplan:1992bt}
\bibitem{Kaplan:1992bt}
  D.~B.~Kaplan,
  ``A Method for simulating chiral fermions on the lattice,''
  Phys.\ Lett.\  B {\bf 288}, 342 (1992)
  [hep-lat/9206013].
  %%CITATION = PHLTA,B288,342;%%

%\cite{Neuberger:1997fp}
\bibitem{Neuberger:1997fp}
  H.~Neuberger,
  ``Exactly massless quarks on the lattice,''
  Phys.\ Lett.\  B {\bf 417}, 141 (1998)
  [hep-lat/9707022].
  %%CITATION = PHLTA,B417,141;%%

%\cite{Ginsparg:1981bj}
\bibitem{Ginsparg:1981bj}
  P.~H.~Ginsparg and K.~G.~Wilson,
  ``A Remnant Of Chiral Symmetry On The Lattice,''
  Phys.\ Rev.\  D {\bf 25}, 2649 (1982).
  %%CITATION = PHRVA,D25,2649;%%

%\cite{Schubert:2001he}
\bibitem{Schubert:2001he}
  C.~Schubert,
  ``Perturbative quantum field theory in the string-inspired formalism,''
  Phys.\ Rept.\  {\bf 355}, 73 (2001)
  [hep-th/0101036].
  %%CITATION = PRPLC,355,73;%%

  \bibitem{Strassler:1992zr}
  M.~J.~Strassler,
  ``Field theory without Feynman diagrams: One loop effective actions,''
  Nucl.\ Phys.\  B {\bf 385}, 145 (1992)
  [arXiv:hep-ph/9205205].
  %%CITATION = NUPHA,B385,145;%%


%\cite{Gies:2001zp}
\bibitem{Gies:2001zp}
  H.~Gies and K.~Langfeld,
  ``Quantum diffusion of magnetic fields in a numerical worldline approach,''
  Nucl.\ Phys.\  B {\bf 613}, 353 (2001)
  [hep-ph/0102185]; %.
  %%CITATION = NUPHA,B613,353;%%
%
%\cite{Gies:2001tj}
%\bibitem{Gies:2001tj}
%  H.~Gies and K.~Langfeld,
  ``Loops and loop clouds: A numerical approach to the worldline formalism  in
  QED,''
  Int.\ J.\ Mod.\ Phys.\  A {\bf 17}, 966 (2002)
  [hep-ph/0112198].
  %%CITATION = IMPAE,A17,966;%%

%\cite{Gies:2003cv}
\bibitem{Gies:2003cv}
  H.~Gies, K.~Langfeld and L.~Moyaerts,
  ``Casimir effect on the worldline,''
  JHEP {\bf 0306}, 018 (2003)
  [hep-th/0303264].
  %%CITATION = JHEPA,0306,018;%%

%\cite{Gies:2006bt}
\bibitem{Gies:2006bt}
  H.~Gies and K.~Klingmuller,
  ``Casimir effect for curved geometries: PFA validity limits,''
  Phys.\ Rev.\ Lett.\  {\bf 96}, 220401 (2006)
  [quant-ph/0601094]; %.
  %%CITATION = PRLTA,96,220401;%%
%
%\cite{Gies:2006cq}
%\bibitem{Gies:2006cq}
%  H.~Gies and K.~Klingmuller,
  ``Worldline algorithms for Casimir configurations,''
  Phys.\ Rev.\  D {\bf 74}, 045002 (2006)
  [quant-ph/0605141]; %.
  %%CITATION = PHRVA,D74,045002;%%
%
%\cite{Gies:2006xe}
%\bibitem{Gies:2006xe}
%  H.~Gies and K.~Klingmuller,
  ``Casimir edge effects,''
  Phys.\ Rev.\ Lett.\  {\bf 97}, 220405 (2006)
  [quant-ph/0606235].
  %%CITATION = PRLTA,97,220405;%%


%\cite{Langfeld:2002vy}
\bibitem{Langfeld:2002vy}
  K.~Langfeld, L.~Moyaerts and H.~Gies,
  ``Fermion-induced quantum action of vortex systems,''
  Nucl.\ Phys.\  B {\bf 646}, 158 (2002)
  [hep-th/0205304].
  %%CITATION = NUPHA,B646,158;%%



%\cite{Dunne:2005sx}
\bibitem{Dunne:2005sx}
  G.~V.~Dunne and C.~Schubert,
  ``Worldline instantons and pair production in inhomogeneous fields,''
  Phys.\ Rev.\  D {\bf 72}, 105004 (2005)
  [hep-th/0507174];
  %%CITATION = PHRVA,D72,105004;%%
%
%\cite{Dunne:2006st}
%\bibitem{Dunne:2006st}
  G.~V.~Dunne, Q.~h.~Wang, H.~Gies and C.~Schubert,
  ``Worldline instantons. II: The fluctuation prefactor,''
  Phys.\ Rev.\  D {\bf 73}, 065028 (2006)
  [hep-th/0602176].
  %%CITATION = PHRVA,D73,065028;%%
%
%\cite{Gies:2005bz}


\bibitem{Gies:2005bz}
  H.~Gies and K.~Klingmuller,
  ``Pair production in inhomogeneous fields,''
  Phys.\ Rev.\  D {\bf 72}, 065001 (2005)
  [hep-ph/0505099].
  %%CITATION = PHRVA,D72,065001;%%


%\cite{Schmidt:2003bf}
\bibitem{Schmidt:2003bf}
  M.~G.~Schmidt and I.~Stamatescu,
  ``Matter Determinants In Background Fields Using Random Walk World Line Loops
  On The Lattice,''
  Mod.\ Phys.\ Lett.\  A {\bf 18}, 1499 (2003).
  %%CITATION = MPLAE,A18,1499;%%

%\cite{Gross:1974jv}
\bibitem{Gross:1974jv}
  D.~J.~Gross and A.~Neveu,
  ``Dynamical Symmetry Breaking In Asymptotically Free Field Theories,''
  Phys.\ Rev.\  D {\bf 10}, 3235 (1974).
  %%CITATION = PHRVA,D10,3235;%%


   \bibitem{dhn}
  R.~F.~Dashen, B.~Hasslacher and A.~Neveu,
  ``Semiclassical Bound States In An Asymptotically Free Theory,''
  Phys.\ Rev.\  D {\bf 12}, 2443 (1975).
  %%CITATION = PHRVA,D12,2443;%%

%\cite{Gies:2005sb}
\bibitem{Gies:2005sb}
  H.~Gies, J.~Sanchez-Guillen and R.~A.~Vazquez,
  ``Quantum effective actions from nonperturbative worldline dynamics,''
  JHEP {\bf 0508}, 067 (2005)
  [arXiv:hep-th/0505275].
  %%CITATION = JHEPA,0508,067;%%


%\cite{Gawedzki:1985ed}
\bibitem{Gawedzki:1985ed}
  K.~Gawedzki and A.~Kupiainen,
  ``Renormalizing The Nonrenormalizable,''
  Phys.\ Rev.\ Lett.\  {\bf 55}, 363 (1985).
  %%CITATION = PRLTA,55,363;%%


%\cite{Langfeld:2007wh}
\bibitem{Langfeld:2007wh}
  K.~Langfeld, G.~Dunne, H.~Gies and K.~Klingmuller,
  ``Worldline Approach to Chiral Fermions,''
  PoS {\bf LATTICE2007}, 202 (2006)
  [arXiv:0709.4595 [hep-lat]].
  %%CITATION = POSCI,LATTICE2007,202;%%

  \bibitem{as}
  M.~Abramowitz and I.~Stegun, {\it Handbook of Mathematical Functions} (Dover, NY, 1970).

  \bibitem{ww}
  E.~Whittaker and G.~Watson, {\it Modern Analysis}, (Cambridge University Press, 1902).


 \bibitem{brazovskii}
 S.~A.~Brazovskii, S.~A.~Gordynin, and N.~N.~Kirova,
``Exact solution of the Peierls model with an arbitrary number of electrons in the unit cell'',
Pis. Zh. Eksp. Teor. Fiz. {\bf 31}, 486 (1980) [JETP Lett. {\bf 31}, 456 (1980)];
``Excitons, polarons and bipolarons in conducting polymers'', Pis. Zh. Eksp. Teor. Fiz. {\bf 33}, 6 (1981) [JETP Lett. {\bf 33}, 4 (1981)];
S.~A.~Brazovskii, N.~N.~Kirova and Matveenko,
``Peierls effect in conducting polymers'',
Zh. Eksp. Teor. Fiz. {\bf 86}, 743 (1984)
Sov.\ Phys.\ JETP {\bf 59}, 434 (1984).

\bibitem{horovitz}
B.~Horovitz,
``Soliton Lattice in Polyacetylene, Spin-Peierls Systems, and Two-Dimensional Sine-Gordon Systems'',
Phys.\ Rev.\ Lett.\ {\bf 46}, 742  (1981).

\bibitem{mertsching}
J.~Mertsching and H.~J.~Fischbeck,
``The incommensurate Peierls phase of the quasi-dimensional Frohlich model with a nearly half-filled band'', Phys. Stat. Sol. B {\bf 103}, 783 (1981).

\bibitem{buzdin}
A.~I.Buzdin and V.~V.~Tugushev,
``Phase diagrams of electronic and superconducting transitions to soliton lattice states'',
Zh. Eksp. Teor. Fiz. {\bf 85}, 735 (1983), [Sov. Phys. JETP {\bf 58}, 428 (1983)];
K.~Machida and H.~Nakanishi,
``Superconductivity under a ferromagnetic molecular field'',
Phys.\ Rev.\ B {\bf 30}, 122  (1984).


\bibitem{novikov}
S.~P.~Novikov, ``A periodic problem for the Korteweg-de Vries equations'',  Funkts. Anal. i Prilozhen. {\bf 8}, 54 (1974); B.~A.~Dubrovin and S.~P.~Novikov, ``Periodic and conditionally periodic analogues of the many soliton solutions of the Korteweg-de Vries equations,
Zh. Eksper. Teoret. Fiz. {\bf 67}, 2131 (1974).

%\cite{Dunne:1997ia}
\bibitem{Dunne:1997ia}
  G.~V.~Dunne and J.~Feinberg,
  ``Self-isospectral periodic potentials and supersymmetric quantum mechanics,''
  Phys.\ Rev.\  D {\bf 57}, 1271 (1998)
  [arXiv:hep-th/9706012].
  %%CITATION = PHRVA,D57,1271;%%




%\cite{Gies:2008zz}
\bibitem{Gies:2008zz}
  K.~Klingmuller and H.~Gies,
  %``Geothermal Casimir Phenomena,''
  J.\ Phys.\ A  {\bf 41}, 164042 (2008)
  [arXiv:0710.4473 [quant-ph]].
  %%CITATION = JPAGB,A41,164042;%%




%\cite{Wolff:1985av}
\bibitem{Wolff:1985av}
  U.~Wolff,
  ``The Phase Diagram Of The Infinite N Gross-Neveu Model At Finite Temperature
  And Chemical Potential,''
  Phys.\ Lett.\  B {\bf 157}, 303 (1985).
  %%CITATION = PHLTA,B157,303;%%



\bibitem{Kapusta:1989tk}
  J.~I.~Kapusta,
  {\it  Finite Temperature Field Theory}, (Cambridge Univ. Press, 1989).






\bibitem{joshua}
  J.~Feinberg,
  ``All about the static fermion bags in the Gross-Neveu model,''
  Annals Phys.\  {\bf 309}, 166 (2004)
  [arXiv:hep-th/0305240].
  %%CITATION = APNYA,309,166;%%


%\cite{Gilkey:1995mj}
\bibitem{Gilkey:1995mj}
  P.~B.~Gilkey,
 {\it Invariance theory, the heat equation and the Atiyah-Singer index theorem},
(CRC Press, Boca Raton, 1995).
%\href{http://www.slac.stanford.edu/spires/find/hep/www?irn=3625788}{SPIRES entry}

%\cite{Kirsten:2001wz}
\bibitem{Kirsten:2001wz}
  K.~Kirsten,
  {\it Spectral functions in mathematics and physics},
%\href{http://www.slac.stanford.edu/spires/find/hep/www?irn=4785843}{SPIRES entry}
(Chapman and Hall/CRC, Boca Raton,  2001).

%\cite{Vassilevich:2003xt}
\bibitem{Vassilevich:2003xt}
  D.~V.~Vassilevich,
  ``Heat kernel expansion: User's manual,''
  Phys.\ Rept.\  {\bf 388}, 279 (2003)
  [arXiv:hep-th/0306138].
  %%CITATION = PRPLC,388,279;%%



%\bibitem{gattringer}
%  C.~Gattringer, V.~Hermann and M.~Limmer,
%  ``Fermion loop simulation of the lattice Gross-Neveu model,''
%  Phys.\ Rev.\  D {\bf 76}, 014503 (2007)
%  [arXiv:0704.2277 [hep-lat]].
%  %%CITATION = PHRVA,D76,014503;%%




\end{thebibliography}
\end{document}